\documentclass[notoc]{JHEP3}

\usepackage{graphicx}
\usepackage{amsmath,amsfonts}
\usepackage{bm}
\usepackage{textcomp}
\usepackage{color}
\usepackage{array}
\usepackage{verbatim}
\let\normalcolor\relax 

\usepackage[normalem]{ulem}

\long\def\symbolfootnote[#1]#2{\begingroup%
\def\thefootnote{\fnsymbol{footnote}}\footnote[#1]{#2}\endgroup}

\def\hermesauthor[#1]#2{{#2}$^{\, #1}$}
\def\hermesinstitute[#1]#2{$^{#1\,}$ {#2}\\}

\renewcommand{\thefootnote}{\alph{footnote}}
\def\nowat[#1]#2{\(^,\)\footnote[#1]{#2}}

\def\bml{\begin{multline}}
\def\eml{\end{multline}}
\def\x{N^X}            

\def\XU{\left\{\x_u:A^m_{||}\x_u\right\}}

\def\PPP{\csname PPP\endcsname}
\expandafter\def\csname PPP\endcsname{\thetable}

\newcommand{\be}{\begin{equation}}
\newcommand{\ee}{\end{equation}}
\newcommand{\beqa}{\begin{eqnarray}}
\newcommand{\eeqa}{\end{eqnarray}}
\newcommand{\nn}{\nonumber}

\definecolor{lightred}{rgb}{1,0.75,0.75}
\definecolor{lightgreen}{rgb}{0.8,1,0.8}
\definecolor{grey}{rgb}{0.7,0.7,0.7}
\definecolor{darkgreen}{rgb}{0.2,0.6,0.2}

\graphicspath{{./figs/}{./xfigs/}}
\title{
 Inclusive Measurements of Inelastic Electron and Positron Scattering 
from Unpolarized Hydrogen and Deuterium Targets
 \\
}

\author{The HERMES Collaboration}

\abstract{
Results of inclusive measurements of inelastic electron and positron
 scattering from unpolarized
 protons and deuterons  at the HERMES experiment are presented. 
The structure functions $F_2^p$ and $F_2^d$ are determined using a
 parameterization 
 of existing data for  the longitudinal-to-transverse virtual-photon absorption 
cross-section ratio. 
The HERMES results provide data in the ranges $0.006\leq x\leq 0.9$ and 
0.1~GeV$^2\leq Q^2\leq$ 20 GeV$^2$, covering the transition region between
the perturbative and the non-perturbative regimes of QCD  in a so-far largely
unexplored  kinematic region. They are in agreement with 
 existing world data  in the region of overlap.
The measured cross sections  are used, in combination with
 data from other experiments,  to perform fits to the photon-nucleon
cross section using the functional form of the ALLM model.
The deuteron-to-proton cross-section ratio is also determined.
}

\preprint{DC82, paper tag:F2, \today}


\preprint{\today}

\begin{document}
\section*{The HERMES Collaboration}
{%
\begin{flushleft} 
\bf
\hermesauthor[12,15]{A.~Airapetian},
\hermesauthor[26]{N.~Akopov},
\hermesauthor[5]{Z.~Akopov},
\hermesauthor[6]{E.C.~Aschenauer}\nowat[1]{Now at: Brookhaven National Laboratory, Upton, New York 11772-5000, USA},
\hermesauthor[25]{W.~Augustyniak},
\hermesauthor[26]{R.~Avakian}, 
\hermesauthor[26]{A.~Avetissian},  
\hermesauthor[5]{E.~Avetisyan},
\hermesauthor[18]{S.~Belostotski},
\hermesauthor[10]{N.~Bianchi}, 
\hermesauthor[17,24]{H.P.~Blok}, 
\hermesauthor[5]{A.~Borissov},
\hermesauthor[13]{J.~Bowles},
\hermesauthor[19]{V.~Bryzgalov}, 
\hermesauthor[13]{J.~Burns}, 
\hermesauthor[9]{M.~Capiluppi},  
\hermesauthor[10]{G.P.~Capitani}, 
\hermesauthor[21]{E.~Cisbani},  
\hermesauthor[9]{G.~Ciullo},  
\hermesauthor[9]{M.~Contalbrigo}, 
\hermesauthor[9]{P.F.~Dalpiaz},  
\hermesauthor[5]{W.~Deconinck},
\hermesauthor[2]{R.~De~Leo},  
\hermesauthor[11,5]{L.~De~Nardo},  
\hermesauthor[10]{E.~De~Sanctis},  
\hermesauthor[14,8]{M.~Diefenthaler},
\hermesauthor[10]{P.~Di~Nezza},  
\hermesauthor[12]{M.~D\"uren},  
\hermesauthor[12]{M.~Ehrenfried}, 
\hermesauthor[26]{G.~Elbakian},  
\hermesauthor[4]{F.~Ellinghaus},
\hermesauthor[6]{R.~Fabbri},  
\hermesauthor[10]{A.~Fantoni}, 
\hermesauthor[22]{L.~Felawka},  
\hermesauthor[21]{S.~Frullani},
\hermesauthor[6]{D.~Gabbert}, 
\hermesauthor[19]{G.~Gapienko},  
\hermesauthor[19]{V.~Gapienko},  
\hermesauthor[21]{F.~Garibaldi},
\hermesauthor[5,18,22]{G.~Gavrilov}, 
\hermesauthor[26]{V.~Gharibyan},  
\hermesauthor[5,9]{F.~Giordano},  
\hermesauthor[15]{S.~Gliske},
\hermesauthor[6]{M.~Golembiovskaya}, 
\hermesauthor[10]{C.~Hadjidakis},  
\hermesauthor[5]{M.~Hartig}\nowat[2]{Now at: Institut f\"ur Kernphysik, Universit\"at Frankfurt a.M., 60438 Frankfurt a.M., Germany},
\hermesauthor[10]{D.~Hasch},  
\hermesauthor[13]{G.~Hill},  
\hermesauthor[6]{A.~Hillenbrand}, 
\hermesauthor[13]{M.~Hoek},  
\hermesauthor[5]{Y.~Holler},  
\hermesauthor[6]{I.~Hristova}, 
\hermesauthor[23]{Y.~Imazu},
\hermesauthor[19]{A.~Ivanilov},  
\hermesauthor[1]{H.E.~Jackson},
\hermesauthor[11]{H.S.~Jo},  
\hermesauthor[14,11]{S.~Joosten},
\hermesauthor[13]{R.~Kaiser}\nowat[3]{Present address: International Atomic Energy Agency, A-1400 Vienna, Austria},  
\hermesauthor[26]{G.~Karyan},  
\hermesauthor[13,12]{T.~Keri},  
\hermesauthor[4]{E.~Kinney},
\hermesauthor[18]{A.~Kisselev},  
\hermesauthor[19]{V.~Korotkov},  
\hermesauthor[16]{V.~Kozlov}, 
\hermesauthor[8,18]{P.~Kravchenko},
\hermesauthor[7]{V.G.~Krivokhijine}, 
\hermesauthor[2]{L.~Lagamba},  
\hermesauthor[14]{R.~Lamb},
\hermesauthor[17]{L.~Lapik\'as}, 
\hermesauthor[13]{I.~Lehmann},
\hermesauthor[9]{P.~Lenisa},  
\hermesauthor[14]{L.A.~Linden-Levy},
\hermesauthor[11]{A.~L\'opez~Ruiz},
\hermesauthor[15]{W.~Lorenzon},
\hermesauthor[6]{X.-G.~Lu}, 
\hermesauthor[23]{X.-R.~Lu}, 
\hermesauthor[3]{B.-Q.~Ma},
\hermesauthor[13]{D.~Mahon},
\hermesauthor[14]{N.C.R.~Makins},
\hermesauthor[18]{S.I.~Manaenkov}, 
\hermesauthor[21]{L.~Manfr\'e}, 
\hermesauthor[3]{Y.~Mao},
\hermesauthor[25]{B.~Marianski}, 
\hermesauthor[8,4]{A.~Martinez de la Ossa},
\hermesauthor[26]{H.~Marukyan},  
\hermesauthor[22]{C.A.~Miller}, 
\hermesauthor[23]{Y.~Miyachi},  
\hermesauthor[26]{A.~Movsisyan},  
\hermesauthor[10]{V.~Muccifora}  
\hermesauthor[13]{M.~Murray}, 
\hermesauthor[5,8]{A.~Mussgiller},  
\hermesauthor[2]{E.~Nappi},
\hermesauthor[18]{Y.~Naryshkin},
\hermesauthor[8]{A.~Nass}, 
\hermesauthor[6]{M.~Negodaev}, 
\hermesauthor[6]{W.-D.~Nowak}, 
\hermesauthor[9]{L.L.~Pappalardo},  
\hermesauthor[12]{R.~Perez-Benito},
\hermesauthor[8]{N.~Pickert},  
\hermesauthor[8]{M.~Raithel},  
\hermesauthor[1]{P.E.~Reimer},
\hermesauthor[10]{A.R.~Reolon},  
\hermesauthor[6]{C.~Riedl},
\hermesauthor[8]{K.~Rith},  
\hermesauthor[13]{G.~Rosner}, 
\hermesauthor[5]{A.~Rostomyan},  
\hermesauthor[14]{J.~Rubin},
\hermesauthor[11]{D.~Ryckbosch},  
\hermesauthor[19]{Y.~Salomatin},  
\hermesauthor[23,20]{F.~Sanftl},  
\hermesauthor[20]{A.~Sch\"afer},  
\hermesauthor[6,11]{G.~Schnell}\nowat[4]{Now at: Department of Theoretical 
Physics, University of the Basque Country UPV/EHU, 48080 Bilbao, Spain and
IKERBASQUE, Basque Foundation for Science, 48011 Bilbao, Spain},  
\hermesauthor[5]{K.P.~Sch\"uler},  
\hermesauthor[13]{B.~Seitz},
\hermesauthor[23]{T.-A.~Shibata},  
\hermesauthor[7]{V.~Shutov},  
\hermesauthor[9]{M.~Stancari}, 
\hermesauthor[9]{M.~Statera}, 
\hermesauthor[8]{E.~Steffens},  
\hermesauthor[17]{J.J.M.~Steijger},  
\hermesauthor[12]{H.~Stenzel},
\hermesauthor[6]{J.~Stewart}, 
\hermesauthor[8]{F.~Stinzing},  
\hermesauthor[26]{S.~Taroian},  
\hermesauthor[25]{A.~Trzcinski}, 
\hermesauthor[11]{M.~Tytgat}, 
\hermesauthor[11]{A.~Vandenbroucke},  
\hermesauthor[11]{Y.~Van~Haarlem},  
\hermesauthor[11]{C.~Van~Hulse},  
\hermesauthor[18]{D.~Veretennikov}, 
\hermesauthor[18]{V.~Vikhrov},  
\hermesauthor[2]{I.~Vilardi},
\hermesauthor[8]{C.~Vogel}, 
\hermesauthor[3]{S.~Wang},
\hermesauthor[6,8]{S.~Yaschenko},  
\hermesauthor[3]{H.~Ye},
\hermesauthor[5]{Z.~Ye},  
\hermesauthor[22]{S.~Yen}, 
\hermesauthor[12]{W.~Yu}, 
\hermesauthor[8]{D.~Zeiler},  
\hermesauthor[5]{B.~Zihlmann}, 
\hermesauthor[25]{P.~Zupranski}.
\end{flushleft} 
}
\bigskip
{\it
\begin{flushleft} 
\hermesinstitute[1]{Physics Division, Argonne National Laboratory, Argonne, Illinois 60439-4843, USA}
\hermesinstitute[2]{Istituto Nazionale di Fisica Nucleare, Sezione di Bari, 70124 Bari, Italy}
\hermesinstitute[3]{School of Physics, Peking University, Beijing 100871, China}
\hermesinstitute[4]{Nuclear Physics Laboratory, University of Colorado, Boulder, Colorado 80309-0390, USA}
\hermesinstitute[5]{DESY, 22603 Hamburg, Germany}
\hermesinstitute[6]{DESY, 15738 Zeuthen, Germany}
\hermesinstitute[7]{Joint Institute for Nuclear Research, 141980 Dubna, Russia}
\hermesinstitute[8]{Physikalisches Institut, Universit\"at Erlangen-N\"urnberg, 91058 Erlangen, Germany}
\hermesinstitute[9]{Istituto Nazionale di Fisica Nucleare, Sezione di Ferrara and Dipartimento di Fisica, Universit\`a di Ferrara, 44100 Ferrara, Italy}
\hermesinstitute[10]{Istituto Nazionale di Fisica Nucleare, Laboratori Nazionali di Frascati, 00044 Frascati, Italy}
\hermesinstitute[11]{Department of Subatomic and Radiation Physics, University of Gent, 9000 Gent, Belgium}
\hermesinstitute[12]{Physikalisches Institut, Universit\"at Gie{\ss}en, 35392 Gie{\ss}en, Germany}
\hermesinstitute[13]{Department of Physics and Astronomy, University of Glasgow, Glasgow G12 8QQ, United Kingdom}
\hermesinstitute[14]{Department of Physics, University of Illinois, Urbana, Illinois 61801-3080, USA}
\hermesinstitute[15]{Randall Laboratory of Physics, University of Michigan, Ann Arbor, Michigan 48109-1040, USA }

\hermesinstitute[16]{Lebedev Physical Institute, 117924 Moscow, Russia}

\hermesinstitute[17]{National Institute for Subatomic Physics (Nikhef), 1009 DB Amsterdam, The Netherlands}
\hermesinstitute[18]{Petersburg Nuclear Physics Institute, Gatchina, Leningrad region 188300, Russia}
\hermesinstitute[19]{Institute for High Energy Physics, Protvino, Moscow region 142281, Russia}
\hermesinstitute[20]{Institut f\"ur Theoretische Physik, Universit\"at Regensburg, 93040 Regensburg, Germany}
\hermesinstitute[21]{Istituto Nazionale di Fisica Nucleare, Sezione di Roma, Gruppo Collegato Sanit\`a and Istituto Superiore di Sanit\`a, 00161 Roma, Italy}
\hermesinstitute[22]{TRIUMF, Vancouver, British Columbia V6T 2A3, Canada}
\hermesinstitute[23]{Department of Physics, Tokyo Institute of Technology, Tokyo 152, Japan}
\hermesinstitute[24]{Department of Physics, VU University, 1081 HV Amsterdam, The Netherlands}
\hermesinstitute[25]{Andrzej Soltan Institute for Nuclear Studies, 00-689 Warsaw, Poland}
\hermesinstitute[26]{Yerevan Physics Institute, 375036 Yerevan, Armenia}
\end{flushleft} 
}

\newpage


\section{\label{sec:intro}Introduction}


Over the past decades, 
  lepton-nucleon scattering has played a major role in the development of 
  our present understanding of nucleon structure. For a review on the
  subject see for  example~\cite{sarkar2004}.   
  In lowest order perturbation theory, scattering of charged leptons $l$ off
nucleons $N$ proceeds via the exchange of a neutral boson 
  ($\gamma^*$, $Z^0$). At the HERMES lepton-nucleon centre-of-mass energy of 
$\sqrt{s} = 7.2$ GeV, contributions
  from $Z^0$-exchange to the cross section can be 
  neglected. Therefore, only the electromagnetic interaction in the
  approximation of one-photon exchange is considered here. In
  this approximation, the differential cross section of unpolarized inclusive 
charged-lepton-nucleon
scattering, $l + N \rightarrow l' + X$ (where $X$ denotes the 
undetected final state), is
  parameterized by two structure functions $F_1(x,Q^2)$ and
  $F_2(x,Q^2)$. Here $x = Q^2/2M\nu$ is the Bjorken variable, with $-Q^2$ 
being the square of the four-momentum transferred by the virtual photon and 
$\nu$ its energy in the target rest frame.
The variable $x$ is 
a measure for the inelasticity of the process with $0 \leq x \leq 1$, and 
$x = 1$ for elastic scattering.
 
In the deep-inelastic scattering (DIS) regime, $\sqrt{Q^2}$ and $\nu$ 
 are much larger than the typical hadronic scale, 
usually set to be the mass  $M$ of the nucleon, 
and the invariant mass $W$ of the photon-nucleon system 
is much larger than the masses of nucleon resonances. 
In the Quark-Parton Model (QPM), the DIS process is viewed as 
 the incoherent superposition of elastic 
lepton scattering from quasi-free point-like quarks of any flavor $q$. 
The variable $x$ can then be interpreted as the fraction of the 
longitudinal nucleon momentum carried by the 
struck quark in a frame where the nucleon moves with infinite momentum in the 
direction opposite to that of the virtual photon. 
In this picture, quark distribution functions $f_q(x,Q^2)$ describe the number 
density
of quarks of flavor $q$ in a fast-moving nucleon at a given value of $(x,Q^2)$ 
and experimental values of $F_2(x,Q^2)$ have been used to constrain these.
At low values of $Q^2$, where this picture of incoherent quasi-free scattering 
does not apply, phenomenological models have been developed 
(see e.g. Refs.~\cite{Badelek:1994fe} and~\cite{allm:1991})
to describe the measured structure functions.

\bigskip

A wealth of unpolarized inclusive charged-lepton DIS data is  available from the 
collider experiments H1~\cite{Aid:1996au,Adloff:1997mf,Adloff:1999ah,
Adloff:2000qj,Adloff:2003uh} and 
ZEUS~\cite{zeuslowx,Derrick:1995ef,Derrick:1996hn,Breitweg:1997hz,
Breitweg:2000yn,Chekanov:2001qu} 
at HERA with lepton-nucleon centre-of-mass 
energies $\sqrt{s}$ up to 320 GeV, the muon experiments BCDMS~\cite{bcdms}, EMC~\cite{emceffect}, 
NMC~\cite{nmc9610231} and 
E665~\cite{e665} ($\sqrt{s} \cong  12-31$ GeV), 
experiments with electrons at SLAC~\cite{whitlow:thesis} ($\sqrt{s} \leq  7$~GeV) 
and at JLAB~\cite{Osipenko:2003ua,Osipenko:2005gt,Malace:2009kw,Tvaskis:2010as}) 
($\sqrt{s} \leq  3.25$ GeV).
The HERMES experiment~\cite{Ackerstaff:1998av} at HERA   collected a large data set for 
positron and electron scattering on a variety of 
nuclear targets, including the proton and deuteron data presented here. 
In particular, the HERMES data cover the transition region between
the perturbative and non-perturbative regimes of QCD in a  kinematic region so 
far largely unexplored by other experiments.
In this work, 
these data are presented together with fits to the world data for 
the  photon-nucleon cross section using the Regge-motivated
 approach of the ALLM~\cite{allm:1991,allm:1997} model. 
The paper is organized as follows: 
the formalism leading to the extraction of the structure function 
$F_2$ is briefly reviewed in Sect.~\ref{sec:formalism}; 
Sect.~\ref{sec:hermes} 
deals with the HERMES experimental
arrangement and the data analysis is described in Sect.~\ref{sec:extraction}.
The systematic  uncertainties in the resulting cross sections 
and cross-section ratios  are discussed in Sect.~\ref{sec:sys}. 
Section~\ref{sec:finalresults} 
offers a discussion of the results and
Sect.~\ref{sec:summary} 
provides a summary. 

\section{\label{sec:formalism}Formalism}

\renewcommand{\baselinestretch}{1.}

\TABLE{
{\renewcommand{\baselinestretch}{1.}\caption{\label{table:kinevar}Kinematic
    variables used in the description of lepton scattering. 
}}
 \begin{tabular}{|l|l|}
\hline
\hline
\parbox{0.40\columnwidth}{\raggedright $ m_l$ }                             & \parbox{0.55\columnwidth}{\raggedright Lepton mass (taken to be negligible)  }\\[4mm]  
\parbox{0.40\columnwidth}{\raggedright $ M$ }                               & \parbox{0.55\columnwidth}{\raggedright Mass of target nucleon                              }\\[4mm]  
\parbox{0.40\columnwidth}{\raggedright $k=(E,\vec{k})$, $k'=(E',\vec{k'})$} & \parbox{0.55\columnwidth}{\raggedright 4--momenta of the initial and final state leptons   }\\[4mm]
\parbox{0.40\columnwidth}{\raggedright $\theta,\; \phi$}                    & \parbox{0.55\columnwidth}{\raggedright Polar and azimuthal angle of the scattered lepton   }\\[4mm]
\parbox{0.40\columnwidth}{\raggedright $P\stackrel{\mathrm{lab}}{=}(M,0)$}  & \parbox{0.55\columnwidth}{\raggedright 4--momentum of the initial target nucleon           }\\[4mm]
\parbox{0.40\columnwidth}{\raggedright $q=k-k'$ }                           & \parbox{0.55\columnwidth}{\raggedright 4--momentum of the virtual photon                   }\\[4mm]
\parbox{0.40\columnwidth}{\raggedright $Q^2=-q^2\stackrel{\mathrm{lab}}{\approx} 4EE'\sin^2\frac{\theta}{2}$ }& \parbox{0.55\columnwidth}{\raggedright Negative squared 4--momentum transfer   }\\[4mm]
\parbox{0.40\columnwidth}{\raggedright $ \displaystyle{\nu=\frac{P\cdot q}{M}\stackrel{\mathrm{lab}}{=} E-E'}$ } & \parbox{0.55\columnwidth}{\raggedright Energy of the virtual photon in the target rest frame }\\[4mm]
\parbox{0.40\columnwidth}{\raggedright $ \displaystyle{x=\frac{Q^2}{2\,P\cdot q}~{=}~\frac{Q^2}{2\,M\nu}}$ }       & \parbox{0.55\columnwidth}{\raggedright Bjorken scaling variable }\\[4mm]
\parbox{0.40\columnwidth}{\raggedright $ \displaystyle{y=\frac{P\cdot q}{P\cdot k}\stackrel{\mathrm{lab}}{=}\frac{\nu}{E}}$ }
& \parbox{0.55\columnwidth}{\raggedright Fractional energy of the  virtual photon}\\[4mm]
\parbox{0.40\columnwidth}{\raggedright $W^2=(P+q)^2=M^2+2M\nu-Q^2$}                 & \parbox{0.55\columnwidth}{\raggedright Squared invariant mass of the photon--nucleon system }\\
\hline
\hline
\end{tabular}
}
\renewcommand{\baselinestretch}{1.}

In the approximation of one-photon exchange, 
the inclusive differential cross section for scattering unpolarized
 charged leptons on unpolarized nucleons can be conveniently parameterized  
in terms of the structure functions $F_1(x,Q^2)$ and $F_2(x,Q^2)$:
\begin{equation}
\label{Eq:sigf1f2}
\frac{d^2\sigma}{dx~dQ^2}=\frac{4\pi\alpha^2_{em}}{Q^4}
~\left[y^2\cdot F_1(x,Q^2)+ \left(\frac{1-y}{x}-\frac{My}{2E}\right)\cdot
 F_2(x,Q^2)\right],  
\end{equation}
where $\alpha_{em}$ is the fine-structure constant and 
all other variables are described in Tab.~\ref{table:kinevar}. 
 The quantities $x$ and $Q^2$ are fully 
determined by the kinematic conditions of the incident and scattered leptons
 and the target nucleon.
This cross section can also be written in terms of longitudinal ($L$) 
and transverse ($T$) virtual-photon contributions
\begin{eqnarray}
\label{Eq:cslt}
\frac{d^2\sigma}{dx~dQ^2}=\Gamma[\sigma_T(x,Q^2)+\epsilon ~\sigma_L(x,Q^2)]~, 
\end{eqnarray}
where  $\sigma_L$ and $\sigma_T$ are the
absorption cross sections for 
longitudinal and transverse virtual photons, $\Gamma$ is the flux of 
transverse virtual photons and the virtual-photon polarization 
parameter $\epsilon$ is the ratio of virtual-photon fluxes for 
longitudinal and transverse polarizations~\cite{close}.
The structure functions $F_1(x,Q^2)$ and $F_2(x,Q^2)$ can then be expressed in 
terms of the two virtual-photon absorption cross sections $\sigma_L(x,Q^2)$ and 
$\sigma_T(x,Q^2)$:
\begin{eqnarray}
\label{Eq:f1lt}
& &F_1(x,Q^2)=\frac{1}{4\pi^2\alpha_{em}}~MK\cdot \sigma_T(x,Q^2)~,\\
& &F_2(x,Q^2)=\frac{1}{4\pi^2\alpha_{em}}~\frac{\nu K}{1+\displaystyle\frac{Q^2}{4M^2x^2}}\cdot\left[\sigma_L(x,Q^2)+\sigma_T(x,Q^2)\right]
~,
\label{Eq:f2lt}
\end{eqnarray}
where $K=\nu(1-x)$ 
in the Hand convention~\cite{handconvention1,handconvention2}.
The longitudinal-to-transverse photon-absorption cross-section ratio 
$R=\sigma_L/\sigma_T$ 
can be expressed in terms of $F_1$ and $F_2$:
\begin{equation}
\label{Eq:rgeneral}
R(x,Q^2)=\frac{\sigma_L}{\sigma_T}=\left(1+\frac{4M^2x^2}{Q^2}\right)
\frac{F_2(x,Q^2)}{2xF_1(x,Q^2)}-1~.
\end{equation}

A determination of the structure functions $F_1(x,Q^2)$ and
 $F_2(x,Q^2)$  requires in principle cross-section measurements 
  made at the same $x$
  and $Q^2$ but at two or more different values of $y$ 
(see Eq.~(\ref{Eq:sigf1f2})), {\it i.e.,} with different beam energies.
The HERMES data used for this analysis were taken at a single beam energy.
In such a situation, it is common practice to re-parameterize
     the cross section, Eq.~(\ref{Eq:sigf1f2}),  as a function of $F_2$ and 
$R$ using Eq.~(\ref{Eq:rgeneral}):
\begin{equation}
\label{Eq:sigf2r}
\frac{d^2\sigma}{dx~dQ^2}=\frac{4\pi\alpha^2_{em}}{Q^4}\frac{F_2(x,Q^2)}{x}
 \left[1-y-\frac{Q^2}{4E^2}+
\frac{y^2+Q^2/E^2}{2[1+R(x,Q^2)]}\right]~.
\end{equation}
The structure function $F_2(x,Q^2)$ can then be extracted from a single 
cross-section measurement at a
 given ($x,Q^2$), by  using a parameterization
for $R(x,Q^2)$ obtained from the available world data. 
This approach has been used in the analysis presented in this paper.


\section{\label{sec:hermes}The Experiment}


The HERA facility at DESY comprised a proton and a lepton storage ring. 
HERMES was a fixed-target experiment using only the lepton beam, 
which consisted of either electrons or positrons at an energy of 27.6 GeV, 
while the proton beam 
passed through the non-instrumented horizontal mid-plane of the HERMES 
spectrometer. 
An open-ended storage cell that could be fed with either polarized or unpolarized 
gas was installed internally to the lepton ring. 

The HERMES spectrometer, which consisted of two identical halves above 
and below the electron beam, was a forward
spectrometer~\cite{Ackerstaff:1998av}
 with multiple tracking stages before and after a 1.5~Tm dipole magnet. 
It had a geometrical acceptance of
$\pm170$~mrad horizontally and $\pm(40-140)$~mrad vertically
for particles originating from the center of the target cell, 
resulting in polar scattering angles $\theta$ ranging from about 40 to 220~mrad.
Particle identification (PID) capabilities were provided by combining the 
responses of a lead-glass calorimeter, a pre-shower hodoscope (H2), a  
transition-radiation detector (TRD), and a  threshold \v{C}erenkov detector 
that was upgraded to a dual-radiator
ring-imaging \v{C}erenkov detector (RICH)~\cite{Akopov:2000qi,rich2}
 in the year 1998.
The lead-glass calorimeter and the pre-shower hodoscope  were included in the
trigger together with two other  hodoscopes (H0 and H1).

 In this experiment target gases of  hydrogen and deuterium were used.
Part of the data were taken  with polarized hydrogen and deuterium, 
with the target spin being
reversed in 1-3~min time intervals so that the target was effectively 
unpolarized.
In the case of hydrogen, using areal densities of the order of 
10$^{14}$~nucleons~cm$^{-2}$
 and lepton currents of typically about 30~mA,  luminosities  of the order of 
$2\cdot 10^{31}$~cm$^{-2}$s$^{-1}$ were achieved for the polarized running, 
and about 10 times higher values for unpolarized running.
The luminosity was measured by scattering  the lepton beam off the atomic 
electrons of the target gas,
{\it i.e.}, M\o ller scattering $e^-  e^- \rightarrow e^-  e^-$ for an 
electron beam and Bhabha scattering $e^+  e^- \rightarrow e^+  e^-$ together
with the annihilation process $e^+  e^- \rightarrow \gamma \gamma$ for a
positron beam. 
The cross sections for these processes are precisely known in Quantum
 Electrodynamics, including radiative corrections.
The scattered particles were detected in coincidence by two identical small
calorimeters~\cite{Benisch:2001rr} located symmetrically with
respect to the beam pipe.
The coincidence rate of the pairs of leptons 
(and photons) provided a relative monitor of the luminosity. An
absolute calibration of the luminosity measurement 
was provided by correlating the coincidence rate
with the yields of the  M\o ller, Bhabha and annihilation processes.


\section{\label{sec:extraction}Data Analysis}


An event is accepted if it contains a track identified  as a lepton by
the PID system (see Sect.~\ref{subs:PID}), and satisfies the selection criteria
     described in Sect.~\ref{sect:kine}.
The number of measured events $N_{meas}$ in each ($x,Q^2$) bin is corrected by 
subtracting the charge-symmetric background from secondary processes $N_{cs}$ 
and by dividing the resulting number by the
corresponding  trigger and lepton-identification efficiencies 
${\cal E}_{trigger}$ and ${\cal E}_{lep}$, 
while taking into account the hadron contamination ${\cal C}_{had}$:
\begin{equation}\label{eq:corrections}
\displaystyle 
N_{events}=(N_{meas}-N_{cs})\cdot \frac{1}{{\cal{E}}_{trigger}}\cdot 
\frac{1-{\cal{C}}_{had}}{{\cal{E}}_{lep}}~.
\end{equation}
These corrections are described in Sects.~\ref{subsec:trigger-eff} to 
\ref{subs:csb}.

The experimental cross section is then obtained as the ratio of the number
 of events $N_{events}$ in each $(x,Q^2)$ bin of widths $\Delta x $ and 
$\Delta Q^2$, and 
the  integrated luminosity $L$ (see Sect.~\ref{subsec:luminosity}):
\begin{equation}
\displaystyle 
\frac{d^2\sigma_{Exp}}{d x ~d Q^2}(x,Q^2)=\frac{N_{events}(x,Q^2)}{
\Delta x ~\Delta Q^2}\cdot \frac{1}{L}~.
\label{Eq:yields}
\end{equation}

An  unfolding procedure for  disentangling instrumental and
radiative effects from the measured cross section 
is then applied in order to obtain the  Born
cross sections $\sigma_{Born}^{p,d}\equiv\sigma^{p,d}$ 
(see Sect.~\ref{subsec:smearing}). 
The structure functions $F_2^p$ and $F_2^d$ are finally derived from
the Born  cross sections  
through Eq.~(\ref{Eq:sigf2r}) using the parameterization 
$R=R_{1998}$~\cite{Abe:R1998}.
Two more corrections related to detector geometry and alignment 
are discussed in Sects.~\ref{subsec:bh} and \ref{subs:misal}.


\subsection{Event selection}\label{sect:kine}


The kinematic range of the events selected for this analysis is shown in
Fig.~\ref{fig:xq2plane}, together with the requirements imposed on the
kinematic and geometrical variables. 
\FIGURE{
\includegraphics[width=0.7\textwidth]{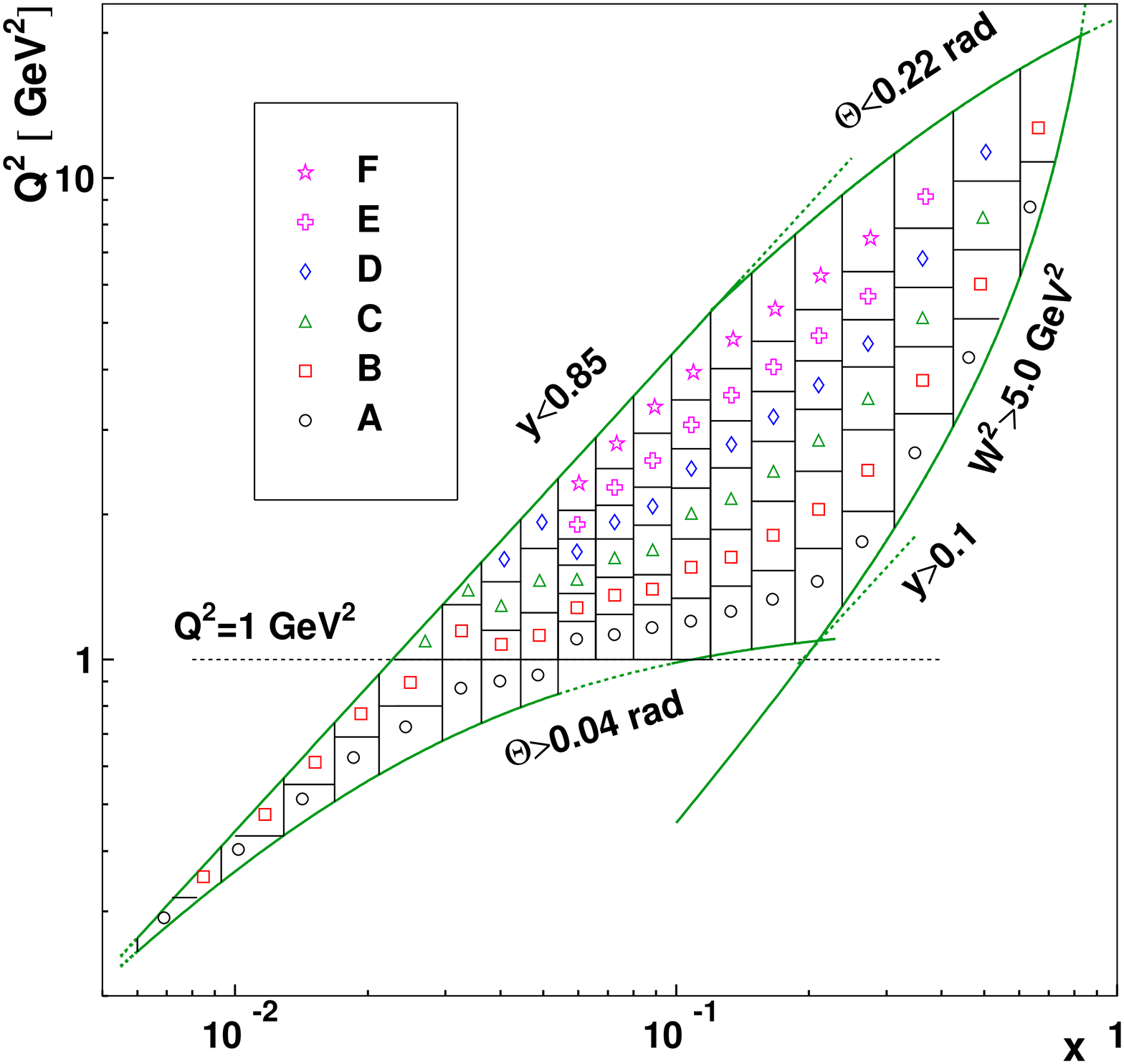}
\caption{Binning in $(x,Q^2)$ used in the analysis and kinematic
    acceptance of events. 
The  kinematic region covered is limited by the geometrical acceptance
in $\theta$ and constraints on $y$ and $W^2$.
The symbols mark the locations of the average values of $(x, Q^2)$
  of each bin.
The symbols A to F denote bins with increasing $Q^2$ at given $x$.
\label{fig:xq2plane}}
}
\TABLE{
\begin{tabular}[h]{|l|r|r|}
\hline 
\hline 
Year &  \multicolumn{2}{|c|}{events, in  million}\\
\hline
 &{\bf proton}&{\bf deuteron}\\
\hline
$1996$  & 2.3  & 2.8 \\
$1997$  & 4.6 & 3.1 \\
$2000$  & 9.5 & 12.6  \\
\hline
total & 16.4 & 18.5\\
\hline
\hline
\end{tabular}
\caption{\label{tab:yields} Number of raw events $N_{meas}$ used in 
this analysis, separated into the years of data taking.
The numbers correspond to 
the total luminosities of about 450\,pb$^{-1}$ on the proton and about 
460\,pb$^{-1}$ on the deuteron.}
}
The tracks are required to be fully contained within 
the fiducial geometric acceptance of the HERMES spectrometer. 
The constraint $W^2>5$~GeV$^2$ 
excludes  the region of nucleon resonances and acts as a selection
for $y>0.1$. 
The constraint $y\leq 0.85$
discards the low-momentum region, where radiative effects increase
and where the trigger efficiency has not yet reached its  plateau.
The requirements imposed on $y$ select the momentum range $4.1$\,GeV$<p<24.8$\,GeV
for the detected particle.
The resulting ($x,Q^2$) region,  $0.006\leq x \leq 0.9$ 
and $0.1$~GeV$^2$~$\leq Q^2\leq 20$~GeV$^2$,   
is subdivided into 19 bins in $x$ and  each $x$ bin into  up to six bins 
in $Q^2$. 

Table~\ref{tab:yields} shows the numbers of events 
for each year of measurements used in this analysis, before the application of 
any of the corrections discussed in the following sections.


\subsection{\label{subsec:trigger-eff}Trigger }


The  trigger  used for the recording of inelastic scattering events required 
signals from the hodoscopes H0, H1, and H2 and a sufficiently large 
energy deposition in the calorimeter. 
The efficiencies of the trigger detectors are extracted individually 
from special calibration triggers and combined 
to obtain the total trigger efficiency.
It is assumed that inefficiencies in the electronic trigger logics 
are negligible. 
Trigger efficiencies are sensitive to, among others, 
misalignment effects,  detector-voltage setting, and radiation damage, 
the latter especially in the H0 hodoscope. 
This last effect is responsible for the reduced efficiencies seen  
at small scattering angles
and for the differences between the top and bottom detector. 
Such differences are  shown in Fig.~\ref{fig:trigeff2000}
for data taken in the year 2000, for the kinematic binning used in the analysis. 
\FIGURE{
\includegraphics[width=0.7\textwidth]{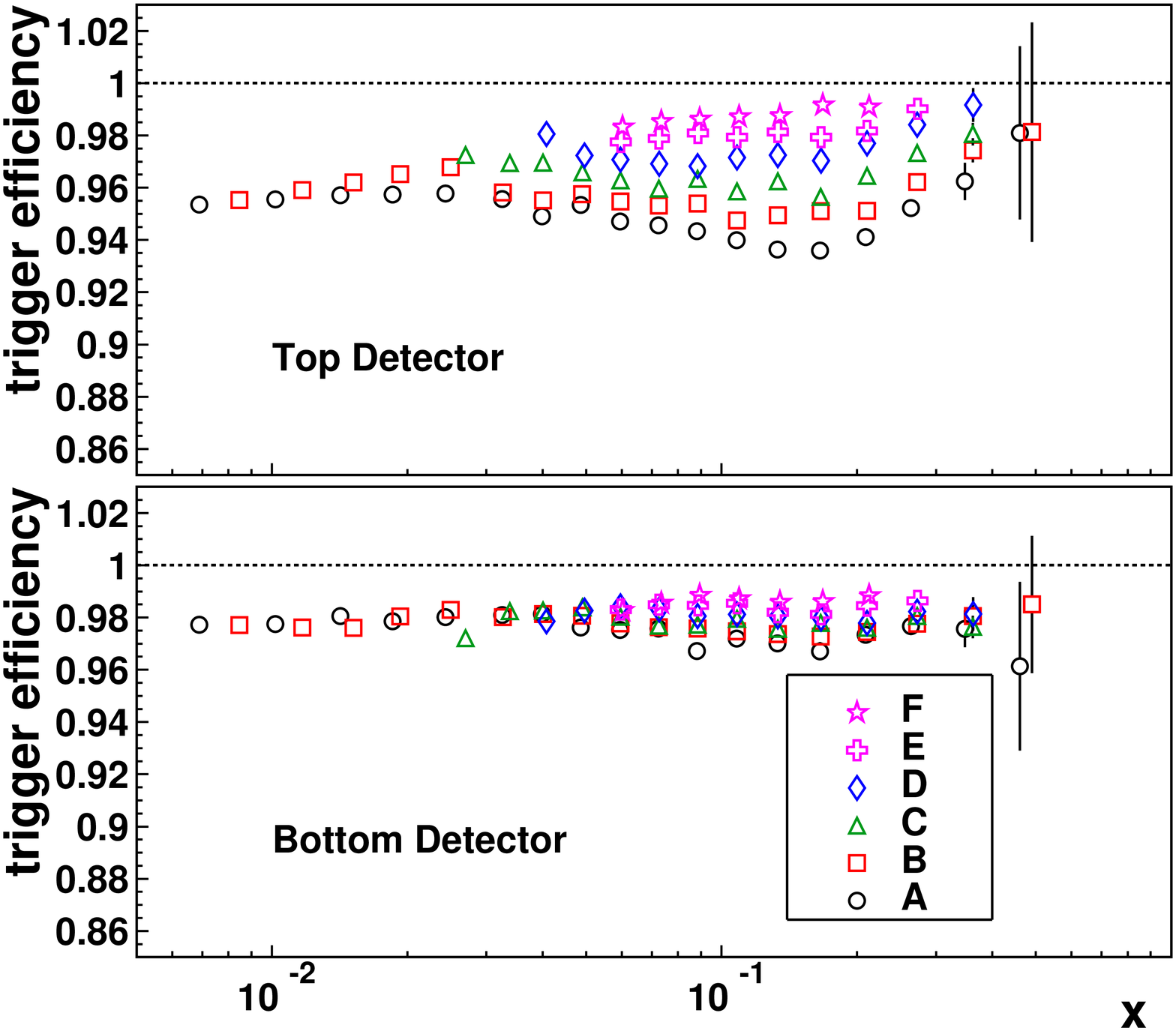}
\caption{
Trigger efficiencies ${\cal{E}}_{trigger}$ for data taken in the year 2000 
shown separately for the top and bottom spectrometer halves. The error
  bars represent only statistical uncertainties. 
The symbols refer to the $Q^2$ bins shown in Fig.~\ref{fig:xq2plane}.}
\label{fig:trigeff2000}
}


\subsection{Particle Identification}\label{subs:PID}


The scattered lepton (positron or electron) is identified by 
a combination of the responses  of the transition-radiation detector TRD, 
the pre-shower hodoscope H2, 
and the lead-glass calorimeter. Each of these elements used alone gives a high
rejection of hadrons. 
A \v{C}erenkov detector provides additional hadron identification.
(A threshold \v{C}erenkov was used for pion identification in 1996-97, and a 
ring-imaging \v{C}erenkov detector was used thereafter to identify pions, kaons,
and protons.) 
The detector response of an individual PID element is determined
by placing very restrictive constraints on the response of the remaining 
elements, thereby generating a clean sample of a given kind of particles 
 with which the unit under study is calibrated.
 
In combination, the array of detectors provides an average lepton
identification with an efficiency ${\cal{E}}_{lep}$ 
of about 98\% and an average hadron contamination ${\cal{C}}_{had}$
below 1\% over the full kinematic range of the HERMES acceptance. 
At low values of $x$, lepton efficiencies as low as 
94\% and hadron contaminations as high as 3\% are reached.

The efficiency for lepton identification and the fractional hadron contamination
as a function of $x$ for the various $Q^2$ bins 
 are presented in Fig.~\ref{fig:pideff2000}
 for representative data taken in the year 2000.
The figure shows  that for smaller values of $x$ ($x<0.1$) a lower
        lepton identification efficiency appears correlated to a larger
        hadron contamination.
\FIGURE{
\includegraphics[width=0.7\textwidth]{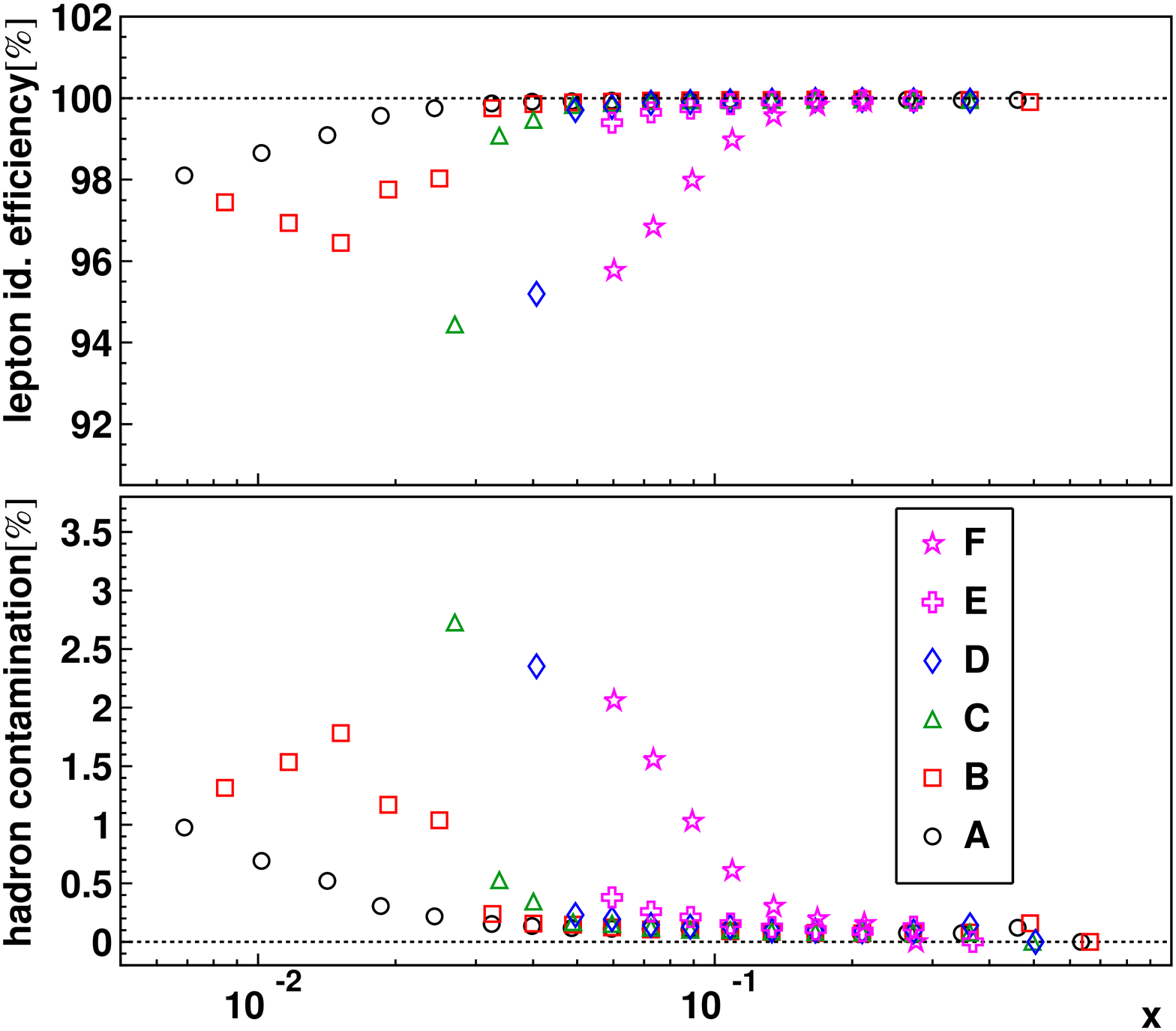}
\caption{
Lepton identification efficiency ${\cal{E}}_{lep}$ and hadron 
contamination ${\cal{C}}_{had}$ 
in the year 2000.  
The symbols refer to the $Q^2$ binning shown in Fig.~\ref{fig:xq2plane}.
\label{fig:pideff2000}}
}


\subsection{Charge-symmetric background}\label{subs:csb}


The observed event sample is  contaminated by 
background coming mostly from charge-symmetric processes, such as 
meson Dalitz decays (e.~g. $\pi^0\rightarrow e^+e^-\gamma$) 
or photon conversions into $e^+e^-$ pairs. 
Since  these positrons and electrons originate from secondary processes, they
 typically have lower momenta and are thus concentrated at high $y$.
A correction for charge-symmetric background events $N_{cs}$
 is applied in each kinematic bin
by counting with negative weight leptons with a charge opposite to that
of the beam particle. It is assumed that acceptance and inefficiencies 
are the same for background electrons and positrons, even though 
their spatial distributions after the magnet are quite 
different.
The $x$ dependence of CS, the ratio of 
charge-symmetric events to the total number of events in each kinematic bin, 
is shown for the six $Q^2$ bins in Fig.~\ref{fig:xq2cs}.
The charge-symmetric background is negligible at large particle momenta, 
but reaches up to 12\% at low particle momenta of about 6~GeV. 


\subsection{\label{subsec:luminosity}Luminosity}


The integrated luminosity $L$ per nucleon is calculated as flollows: 
\begin{equation}
\displaystyle 
L=\int\hspace{-0.085cm}\mathcal{L}\, dt  
=(R_{LR}-2\Delta t \cdot R_{L} \cdot R_{R}) \cdot c_{live}
 \cdot C_{Lumi}\cdot \Delta b \cdot \frac{A}{Z}
 ~.
\label{Eq:luminosity}
\end{equation}
Here, $R_{L}$ and $R_{R}$ are the count rates in the left
and right luminosity detector, respectively, $R_{LR}$ is the coincidence
rate measured within a time window of $\Delta t=40$~ns,
$c_{live}$ is the trigger livetime factor,
$C_{Lumi}$ is the year-dependent luminosity factor, 
$\Delta b$ is the time interval in which the luminosity rates were obtained,
and $A/Z$ is the ratio of the numbers of nucleons ($A$) and 
electrons ($Z$) in the target gas atoms.
\FIGURE{
\includegraphics[width=0.5\columnwidth]{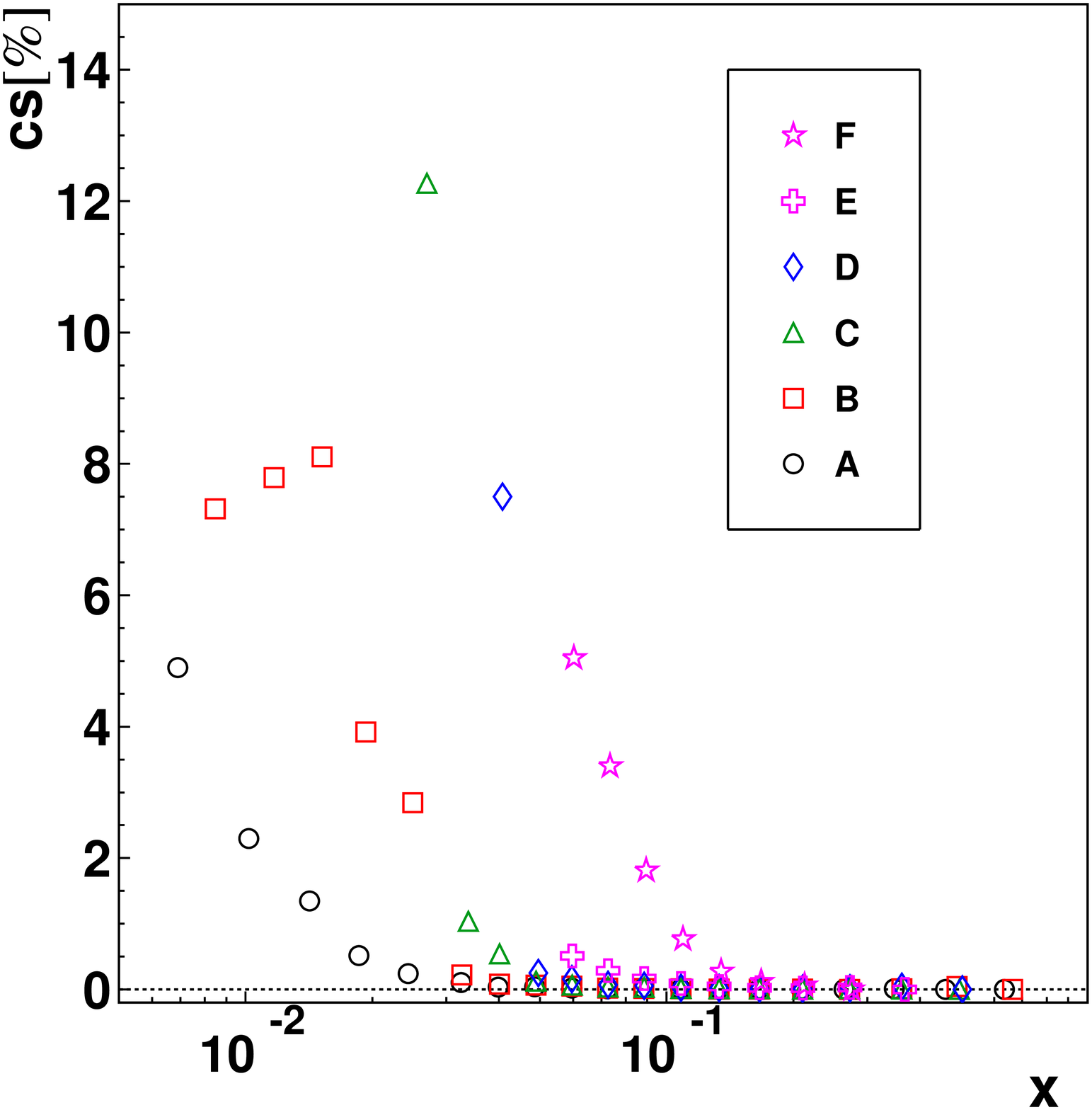}
\caption{Percentage of charge-symmetric background, calculated
 from the ratio of the charge-symmetric events to the total events in each bin, 
for the 2000 deuterium data.
The symbols refer to the $Q^2$ binning shown in Fig.~\ref{fig:xq2plane}.
\label{fig:xq2cs}}
}
The term $2\Delta t \cdot R_{L} \cdot R_{R}$ in Eq.~(\ref{Eq:luminosity})
corrects for accidental coincidences according to the statistical 
expectation and is of the order of 0.1-0.5\%.
The physics trigger livetime contribution $c_{live}$
is defined as the fraction of the physics events that
are accepted by the data acquisition system out of all 
events generating a physics trigger. 
This quantity is typically above $90\%$.

The data aquisition system of the luminosity detector worked
independently of the physics triggers.
It is assumed that the inefficiency of the luminosity event trigger 
was negligible.
The luminosity factor $C_{Lumi}$ accounts for 
the geometric acceptance of the luminosity detector, the beam position
and the absolute M{\o}ller and Bhabha cross sections. 
Its year  dependence derives from the ageing of the luminosity
detector and different running
conditions, {\it i.e.}, changes in  beam charge  and
beam optics.
The dependence of the coincidence rate from beam position and  slope
was measured in order to disentangle the dependence of the measured 
coincidence rate 
from beam orbits and  geometrical acceptance of the luminosity detector.
The uncertainty in the measurement of the absolute luminosity is dominated 
by the uncertainty on the  acceptance of the detector, which 
depends sensitively on the impact coordinates of the particle. 
The uncertainty on the latter is about 2.5~mm, which propagates 
into an uncertainty of about 7\% on the integrated cross section and therefore on 
the luminosity.


\subsection{\label{subsec:smearing}Instrumental smearing and radiative effects}


Instrumental smearing is due to intrinsic detector resolution and 
multiple scattering in the various 
detector elements of the particles emerging from the DIS process 
and identified as the scattered lepton.
Radiative effects include vertex corrections to the
QED hard scattering amplitude and 
radiation of one or more real photons by the incoming or outgoing  lepton.
Radiative effects and instrumental smearing
both modify the Born kinematic conditions
resulting in altered reconstructed kinematic variables. 
Migration probabilities for the relevant kinematic variables are determined 
 from a Monte Carlo simulation and used to correct the measured distributions.  

The Born cross section for inelastic scattering on the proton 
is simulated according to 
the ALLM97 parameterization of $F_2^p$~\cite{allm:1997} and 
the parameterization $R_{1990}$~\cite{r1990} for $R$, 
while that for the deuteron
 is derived from the same parameterizations in conjunction with 
the fit~\cite{nmcratiofit} to $F_2^d/F_2^p$ data from NMC, SLAC and BCDMS.
Radiative effects are simulated with RADGEN~\cite{radgen}. 
The electric and magnetic form factors of the proton and neutron, 
from which the elastic cross sections are derived, are taken 
from the fits in Refs.~\cite{formfactorp} and~\cite{formfactorn}.
When using the more recent parameterizations for the proton 
from Ref.~\cite{Arrington:2007ux}, the results
are essentially the same.

The probabilistic information about event migration 
can be summarized in a smearing matrix~\cite{hermesg1},
\begin{eqnarray}
\label{Eq:smearing}
\displaystyle S(i,j)=\frac{\partial \sigma_{Exp}(i)}{\partial \sigma_{{Born}}(j)}=
\frac{n(i,j)}{n_{Born }(j)}~.
\end{eqnarray} 
Here, $n(i,j)$ is the migration matrix  representing  
the number of events originating from kinematic bin $j$ at Born level and 
measured in bin $i$. 
It is extracted from a Monte Carlo simulation with full track reconstruction that 
simulates the inelastic scattering cross section, QED radiative effects
and  instrumental smearing.
Material outside the detector acceptance is excluded from this
  simulation for computational economy.  
The vector $n_{Born}(j)$ containing the  number of events
 at Born level is 
obtained from a second Monte Carlo calculation that simulates only the
(unradiated) inelastic cross section.
An additional column $j=0$ is defined for 
events that migrate into the acceptance from outside. 
The smearing matrix $S(i,j)$ has the property of being independent from 
the generated cross section within the acceptance.

The inverted squared submatrix $S'(i,j)=S(i,j>0)$ 
relates the measured
distributions to the distributions at Born level:
\begin{equation}
\displaystyle 
\sigma_{Born}(j) =\sum_i S'^{-1}(j,i)\times [\sigma_{Exp}(i)  - 
S(i,0)\sigma_{Born}(0)].
\label{Eq:unfolding}
\end{equation}

The reconstruction of  simulated tracks
uses the same  algorithm as for real data. 
Tracking-related inefficiencies are  taken into account
in the unfolding procedure, assuming that coincident particles
outside the acceptance do not significantly affect the efficiency
and the simulation adequately models the physical processes in the
tracking detectors. 

The Monte Carlo generated data sample was a factor 10 larger 
than the experimental data sample.
The statistical uncertainties of the Monte Carlo data enter mainly via the
simulated experimental count rates in the migration matrices.
A multi-sampling numerical approach is used to propagate these statistical 
uncertainties through the unfolding algorithm.
The statistical uncertainties of the inelastic scattering Born cross section  
coming from the experimental cross section
and those originating from the finite statistical precision of the Monte
Carlo are summed in quadrature to produce the total statistical uncertainty.


\subsection{\label{subsec:bh}
Detection  efficiency of specific radiative events} 


Radiative corrections include cases where the incoming electron
radiates a high-energy photon and then scatters elastically from the nucleon
with negligible momentum transfer. The efficiency to detect such events is
reduced due to the following effect. The radiated
photon is emitted at small angles and has a large probability to hit the
beam pipe, generating an electromagnetic shower that saturates the wire
chambers. This makes the data acquisition system skip the event
as no tracking is possible.
In order to compensate for this omission, the detection efficiency 
for elastic and quasi-elastic radiative events is estimated using
a dedicated Monte Carlo simulation that includes a complete treatment 
of showers in material outside the geometrical acceptance.

The resulting efficiencies ${\cal{E}}_{e.m.}$ 
are significantly less than 100\% in the range $0.01<x<0.1$. 
They show a dip at $x\simeq 0.02$ where, 
in the case of the proton, they reach values as low as 80\% while in the case 
of  the deuteron they are as low as 90\% (60\%) for elastic (quasi-elastic) 
events. They are  applied to the background term $S(i,0)\sigma_{Born}(0)$ 
in order  to  not over-correct for radiative processes that 
are not observed in the spectrometer. More details 
can be found in Refs.~\cite{hermesg1,hermes:erratum}.

\subsection{\label{subs:misal} Misalignment effects}

Imperfect alignment of the two spectrometer halves
and the beam with respect to their ideal positions  is studied in order to 
estimate the impact on the measured cross sections and structure functions.
Misalignment effects cannot be corrected for in the unfolding because they are not
of a stochastic nature.
Rather, they are studied in a Monte Carlo simulation,
 and the fractional change of the Born cross section in each kinematic bin
is obtained from the ratio of unfolded cross sections when
using a MC with an {\it{aligned}} geometry and another with a 
{\it{misaligned}} geometry.
These fractional changes are  used to rescale the experimental 
Born cross sections on a bin-by-bin basis. 
Misalignment effects are most significant 
for small scattering angles and high particle momenta, {\it i.e.},
at small $Q^2$ in each bin of $x$. The correction reaches values as high as 19\% 
in the lowest $Q^2$ bin and decreases to about 3\% in the highest $Q^2$ bin.


\section{Systematic Uncertainties}\label{sec:sys}


\subsection{Inclusive inelastic scattering cross sections}


\noindent{\it Particle identification.}


\noindent
Correlations between PID detectors as described in Sect.\ref{subs:PID} 
 cannot be completely avoided. They may bias the correction for particle 
identification. These effects are covered by the assignment of a 
conservative PID uncertainty of the full size of the correction 
(see Eq.~(\ref{eq:corrections})).
Hadrons are predominantly produced at small momenta. Thus
particle misidentification 
occurs more likely at high $y$, {\it i.e.},
towards higher $Q^2$ in each bin of $x$.
Nevertheless, the uncertainty due to particle identification, $\delta_{PID}$,  
is always smaller than 3\%, because contaminations somewhat compensate 
inefficiencies.

\vspace{0.5cm}


\noindent{\it Instrumental smearing and radiative effects.}


\noindent
In the unfolding procedure an uncertainty can arise from uncertainties
in the formalism to calculate radiative effects and in the model used 
for the cross section outside the acceptance. 
The latter affect our results through the radiative tail. 
The uncertainty, $\delta_{model}$,  was estimated by varying the input elastic and 
inelastic cross sections within their
uncertainties and found to be below 2\%, 
except for a few bins, where it went up to 4.3\% (3.1\%) at maximum 
for the proton (deuteron) case.
This is negligible compared to 
the overall normalization uncertainty of our data of about 7\% (see below).

\vspace{0.5cm}


\noindent{\it Misalignment.}


\noindent
In each bin, half of the deviations of yields obtained in a Monte Carlo 
simulation with estimated geometric misalignments 
from the yields obtained in a Monte Carlo simulation with aligned (ideal) geometry
serve as an estimate of the systematic uncertainty due to misalignment. 
The uncertainty due to misalignment, $\delta_{mis.}$, reaches values of 
up to 5.4\%. 
However,  the majority of the data points has an uncertainty due to
misalignment  that is smaller than 3\%. 

\vspace{0.5cm}


\noindent{\it Dependence on misalignment of the efficiency  
$\mathcal{E}_{e.m.}$ for elastic and quasi-elastic radiative events.}


\noindent 
The efficiency $\mathcal{E}_{e.m.}$ and its dependence on misalignment 
were studied in Monte Carlo simulations.
The assignment of a corresponding systematic uncertainty 
is accomplished by applying the values of $\mathcal{E}_{e.m.}$ extracted from a 
Monte Carlo simulation with   aligned  and   misaligned geometry to the
high-multiplicity radiative events included in the background term
$S(i,0)\sigma_{Born}(0)$. The
difference of the unfolded inelastic scattering Born cross sections 
obtained for these efficiencies is assigned as a
systematic uncertainty, $\delta_{rad.}$,  due to these radiative corrections.

\vspace{0.5cm}


\noindent{\it Overall normalization uncertainty.}


\noindent
The normalization uncertainties of the absolute cross sections and the
structure functions are dominated by the 
uncertainty of the year-dependent luminosity constant $C_{Lumi}$ in 
Eq.~(\ref{Eq:luminosity}).
The uncertainties of the luminosity constants weighted with the sizes of the 
data sets result in an overall normalization uncertainty of 7.6\% for the 
data taken on a hydrogen target and 
7.5\% for the data taken on a deuterium target. 


\subsection{ Inclusive inelastic scattering  cross-section ratio 
$\sigma^d/\sigma^p$}


The cross-section ratio can be  determined with higher precision than
the cross sections themselves due to the cancellation of 
the misalignment uncertainty, the PID uncertainty and, 
to a large extent, the overall normalization uncertainty. The 
remaining overall normalization uncertainy of 1.4\% 
is attributed to variations of the beam conditions between data sets.

The efficiencies $\mathcal{E}_{e.m.}$ for proton and deuteron 
are different~\cite{HERMES:g1p}, and therefore
do not cancel in the proton-to-deuteron cross-section 
ratio. 
The uncertainty $\delta_{rad}$  of the cross-section ratio 
is obtained by propagating, for proton and deuteron cross sections, 
the uncertainties of  efficiencies for high-multiplicity radiative events 
due to misalignment. 
It is found to be  less than 2.5\% in every kinematic bin.


\section{\label{sec:finalresults}Discussion of the Results }


The kinematic conditions of the HERMES inclusive lepton-nucleon scattering 
cross sections presented here overlap those of existing data over 
a large kinematic range.
New information is provided in the region with
$Q^2\lesssim 1$ GeV$^2$ and $15$\,GeV$^2<W^2<45$\,GeV$^2$, 
corresponding to $0.006<x<0.04$. 


\subsection{Structure functions $F_2^p$ and  $F_2^d$}


The differential cross sections 
$d^2\sigma^{p,d}/dxdQ^2$ for inelastic scattering  on the proton and deuteron 
as well as the corresponding structure functions $F_2^p$ and $F_2^d$ are listed in
Tabs.~\ref{tab:f2p} and \ref{tab:f2d}
together with the statistical and systematic uncertainties. 
The statistical uncertainties of the HERMES data
range between 0.4\% and 3.0\%. Almost 80\% of all
data points have a statistical uncertainty smaller than 1\%.
The overall normalization uncertainty of 7.6\% (7.5\%) for the
inelastic scattering cross  
section on the proton (deuteron) and the contribution from misalignment 
are the dominating systematic uncertainties.

The  differential cross sections 
are shown in Figs.~\ref{fig:sigrp0} and \ref{fig:sigrd0}  as a function of 
$Q^2$ in bins of $x$.
The structure functions are  shown in Figs.~\ref{fig:f2p0} and \ref{fig:f2d0}, 
together with the available world data
from fixed target (E665~\cite{e665}, BCDMS~\cite{bcdms},
NMC~\cite{nmc9610231}, SLAC~\cite{whitlow:thesis},
JLAB~\cite{Osipenko:2003ua,Osipenko:2005gt,Malace:2009kw,Tvaskis:2010as})
 and collider experiments  
(H1 and ZEUS).
The data are overlaid with 
new fits to world data, including the data presented here,
of inclusive proton (GD11-P) and deuteron
(GD11-D) cross sections. These functions are  described in 
section~\ref{app:fits}.
\FIGURE{
\includegraphics[width=\textwidth]{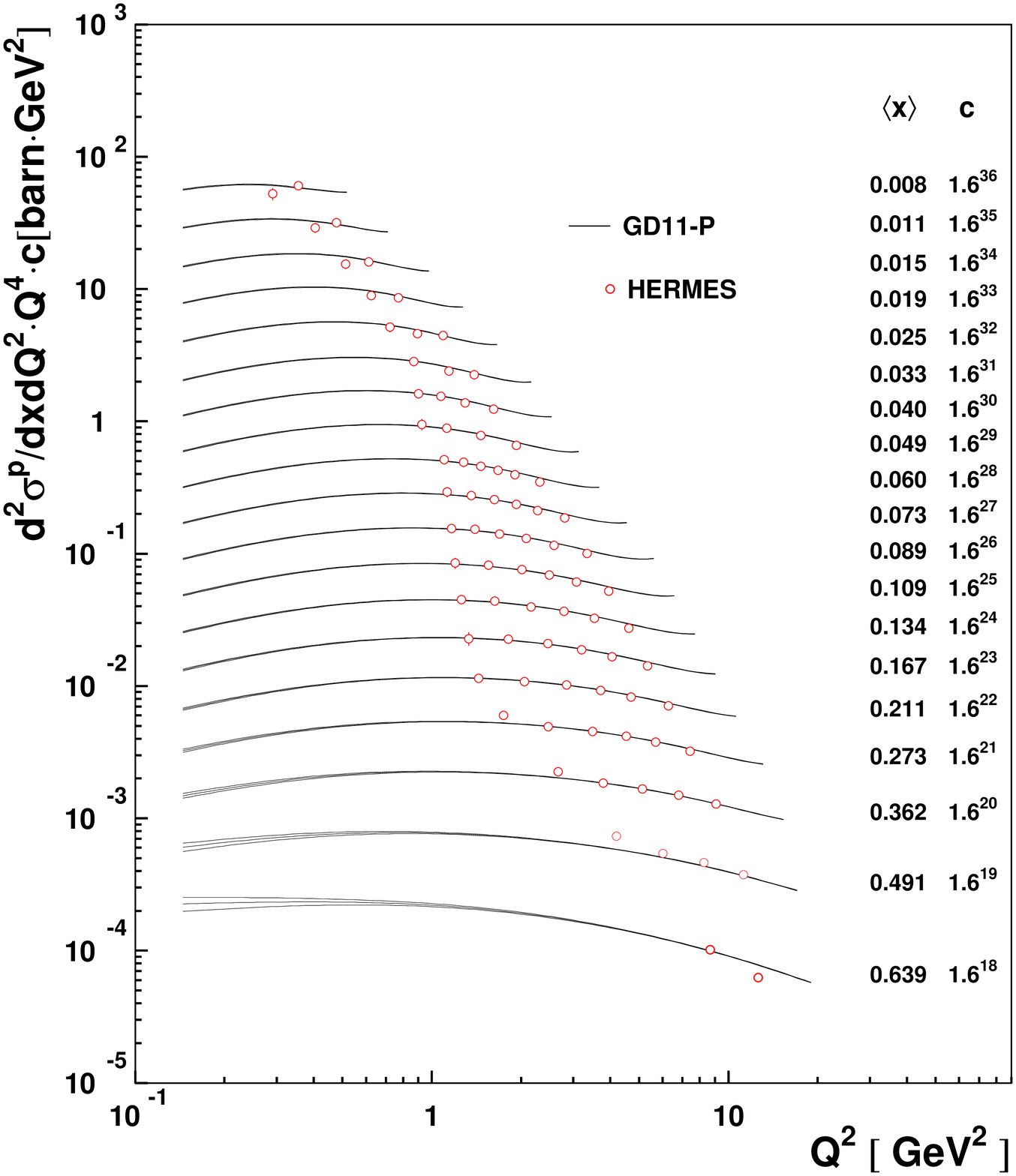}
\caption{\label{fig:sigrp0}
 Inelastic proton 
differential DIS
cross section $d^2\sigma^{p}/dxdQ^2$,
multiplied by a factor $Q^4$ for the purpose of illustration, 
in the kinematic range  $0.008\leq \langle x\rangle \leq 0.639$ and 
$0.2$~GeV$^2\leq \langle Q^2\rangle \leq 20$~GeV$^2$. 
The values of $d^2\sigma^{p}/dxdQ^2\cdot Q^4$ are scaled by powers of $1.6$. 
The results are 
overlaid with the phenomenological parameterization GD11-P 
(central curves) and its
  uncertainty (outer curves). Details on the fits can be found in
  Sect.~\ref{app:fits}. The error bars represent the total uncertainties
calculated as the sum in quadrature of all statistical and systematic 
uncertainties including normalization, and are smaller than the symbols.
}
}
\FIGURE{
\includegraphics[width=\textwidth]{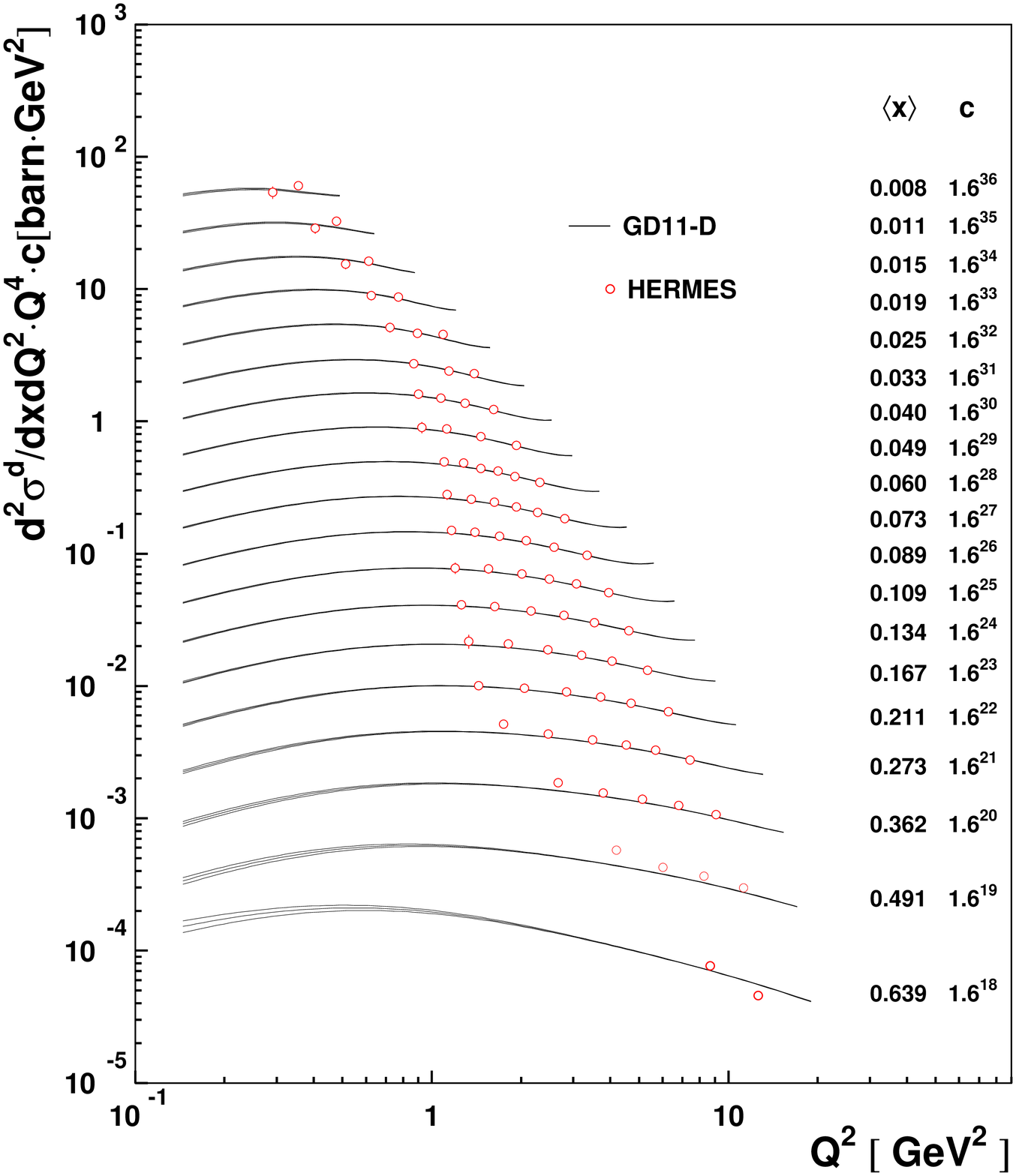}
\caption{\label{fig:sigrd0}
 Inelastic deuteron  differential DIS
cross section $d^2\sigma^{d}/dxdQ^2$,
multiplied by a factor $Q^4$ for the purpose of illustration, 
in the kinematic range  $0.008\leq \langle x\rangle \leq 0.639$ and 
$0.2$~GeV$^2\leq \langle Q^2\rangle \leq 20$~GeV$^2$. 
The values of $d^2\sigma^{d}/dxdQ^2\cdot Q^4$ are scaled by powers of $1.6$. 
The results are 
overlaid with the phenomenological parameterization GD11-D
(central curves) and its
  uncertainty (outer curves). Details on the fits can be found in
  Sect.~\ref{app:fits}. The error bars represent the total uncertainties
calculated as the sum in quadrature of all statistical and systematic 
uncertainties including normalization, and are smaller than the symbols.
}
}
\FIGURE{
\includegraphics[width=0.95\textwidth]{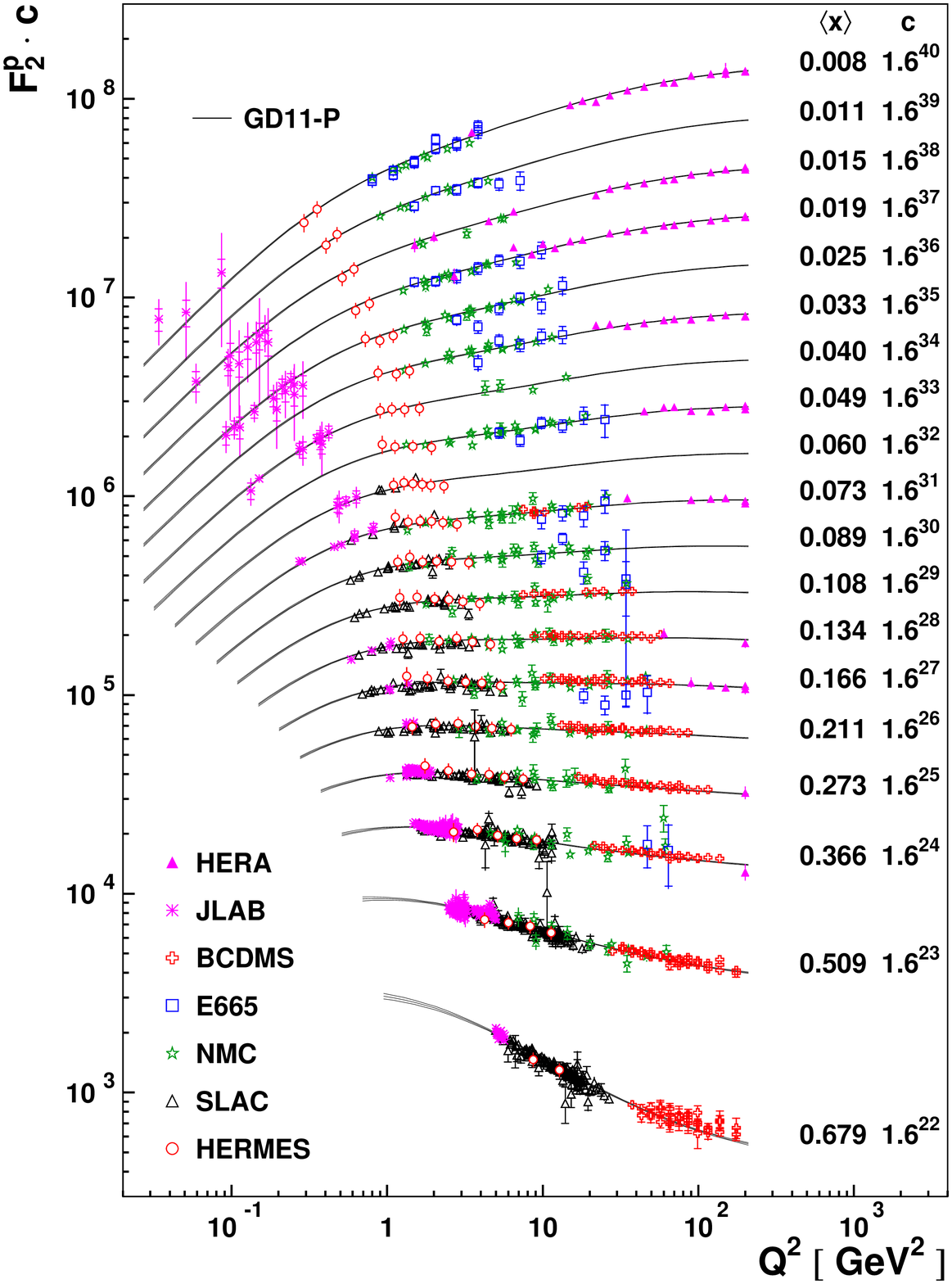}
\caption{\label{fig:f2p0}
HERMES data for $F_2^{p}$ together with world data 
in the kinematic range  $0.008\leq \langle x\rangle\leq 0.679$ and 
$0.02$~GeV$^2\leq \langle Q^2\rangle\leq 20$~GeV$^2$. The results are 
overlaid with the phenomenological parameterization GD11-P
(black solid central curve) and its uncertainty (outer curves) obtained as
  described in Sect.~\ref{app:fits}.
A bin-centering correction is applied to the
 data in order to match the central values of the $x$ bins. 
The values of $F_2^{p}$ are scaled by powers of $1.6$. 
Inner error bars are statistical uncertainties, while
outer error bars are total  uncertainties
calculated as the sum in quadrature of all statistical and systematic 
uncertainties including normalization.
}
}
\FIGURE{
\includegraphics[width=0.95\textwidth]{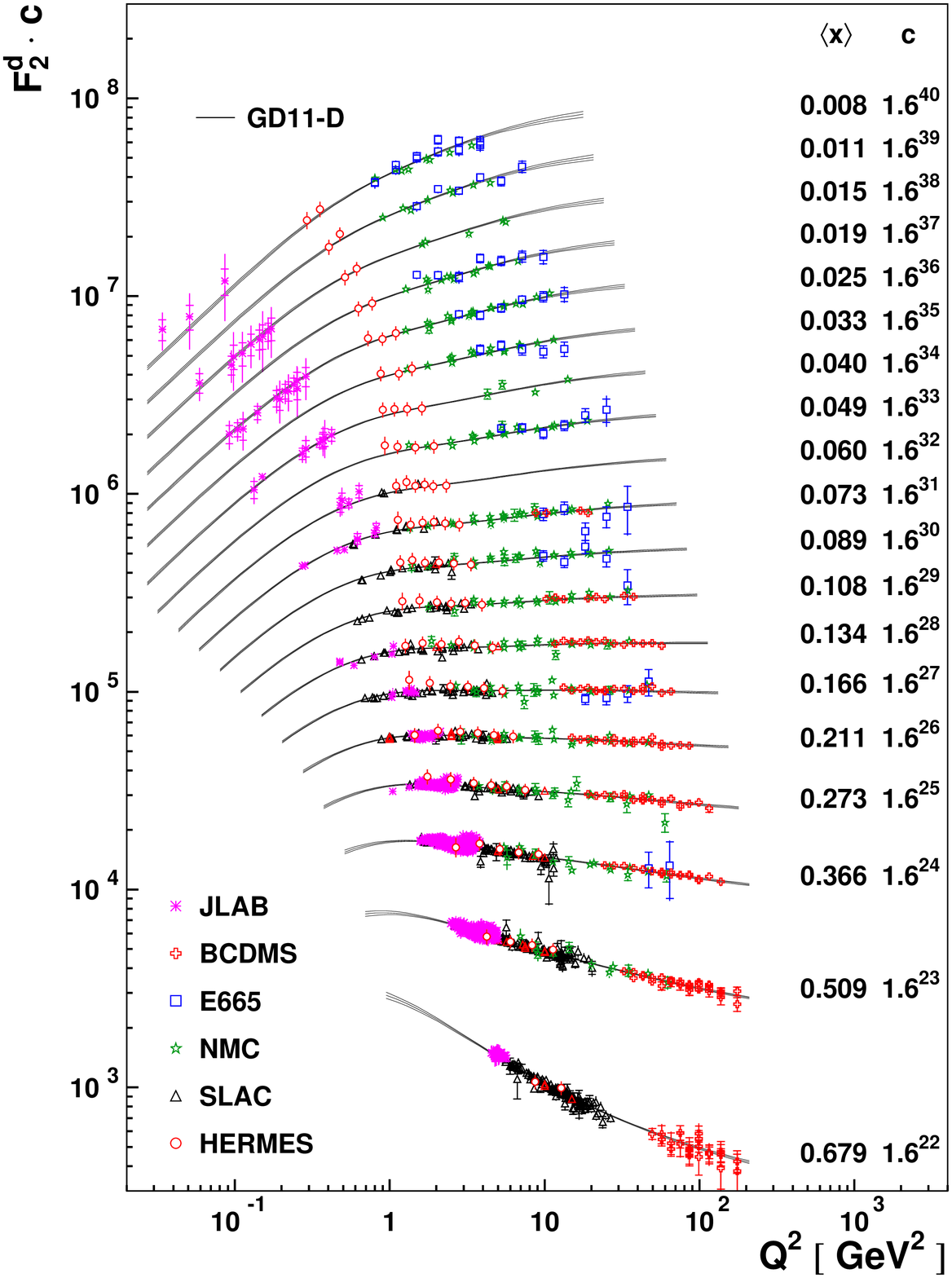}
\caption{\label{fig:f2d0}
HERMES data for $F_2^{d}$ together with world data 
in the kinematic range  $0.008\leq \langle x\rangle\leq 0.679$ and 
$0.02$~GeV$^2\leq \langle Q^2\rangle\leq 20$~GeV$^2$. The results are 
overlaid with the phenomenological parameterization GD11-D
(black solid central curve) and its uncertainty (outer curves) obtained as
  described in Sect.~\ref{app:fits}.
A bin-centering correction is applied to the
 data in order to match the central values of the $x$ bins. 
The values of $F_2^{d}$ are scaled by powers of $1.6$. 
Inner error bars are statistical uncertainties, while
outer error bars are total uncertainties
calculated as the sum in quadrature of all statistical and 
systematic uncertainties including normalization.
}
}

In the region $x\geq 0.07$ and $Q^2>1$~GeV$^2$, HERMES data are in good 
agreement with existing data from SLAC and NMC. 
The HERMES measurement provides also data in a previously 
uncovered kinematic region 
between JLAB data on the one hand and 
 NMC, BCDMS, E665 and the collider experiments on the other.
This can be clearly seen in Figs.~\ref{fig:f2p0} and \ref{fig:f2d0}.

For virtual-photon data ($Q^2>0$), the  photon-nucleon cross section
$\sigma_{L+T}^{p,d}=\sigma_L^{p,d}+\sigma_T^{p,d}$ 
can be derived from the structure function $F_2^{p,d}$ using Eq.~(\ref{Eq:f2lt}):
\begin{equation}
\sigma_{L+T}^{p,d}=4\pi^2 \alpha_{em}\frac{Q^2+4M^2x^2}{Q^4(1-x)}~F_2^{p,d}~.
\end{equation} 
The $W^2$ dependence of the resulting proton and deuteron cross sections
$\sigma_{L+T}^{p,d}$ is shown
in Figs.~\ref{fig:sigltp0} and \ref{fig:sigltd0} 
for HERMES results on 
 inelastic scattering in comparison with  world data and real-photon 
  cross sections.   
HERMES data fill a gap  for $Q^2$ in the range
 0.3-1.~GeV$^2$ and $W^2$ in the range 20-50~GeV$^2$.

\FIGURE{
\includegraphics[width=\textwidth]{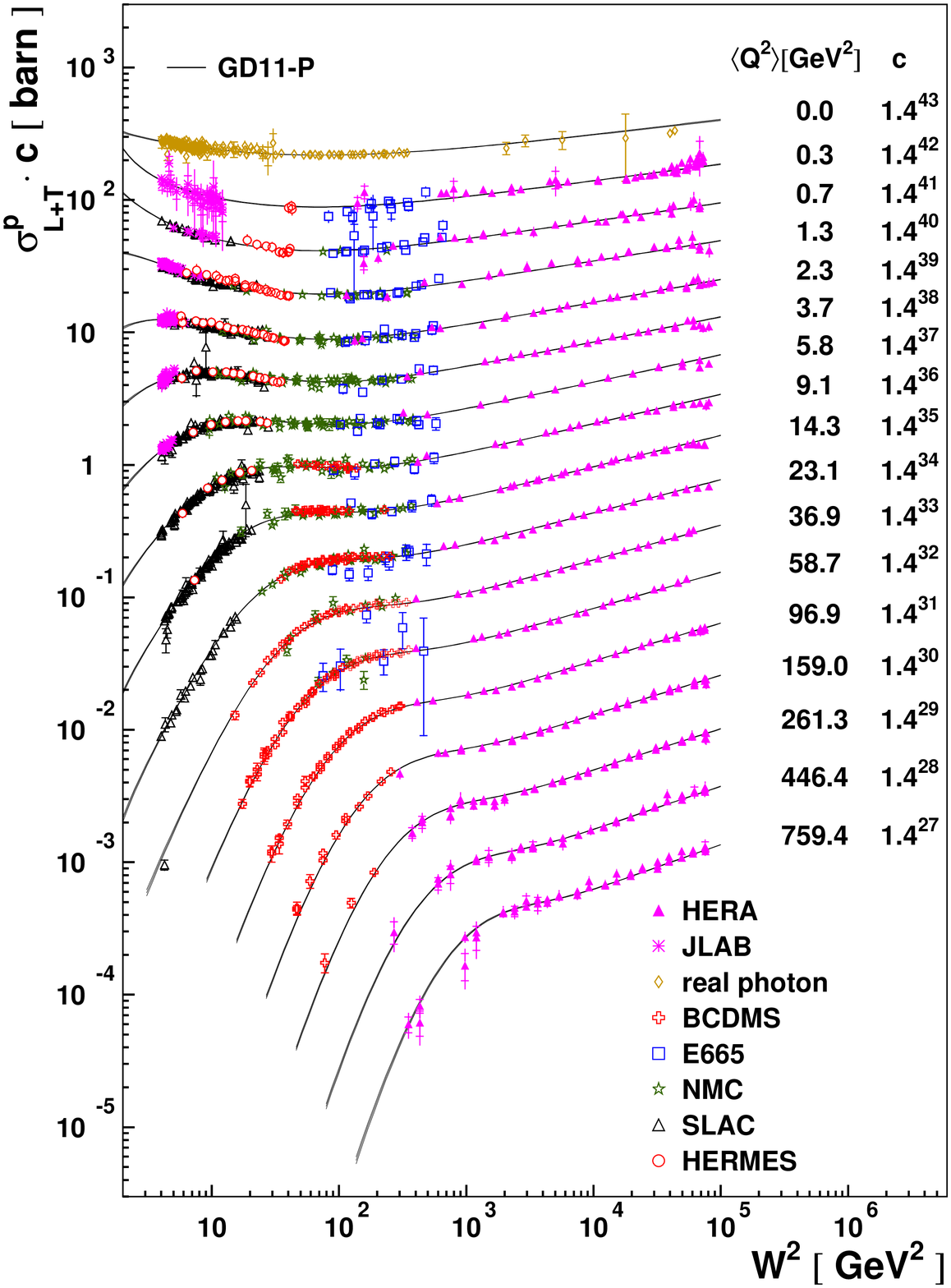}
\caption{\label{fig:sigltp0}
HERMES data for 
the photon-proton 
cross section $\sigma_{L+T}^{p}$  
  as a function of $W^2$, together with world data and the results from
  the GD11-P fit (central curves) and its uncertainties (outer curves), 
in bins of $Q^2$.
  The data points denoted 'real photon' are for
  photoproduction.
Inner error bars are statistical uncertainties, while
outer error bars are total uncertainties
calculated as the sum in quadrature of all statistical and 
systematic uncertainties including normalization.
}
}

\FIGURE{
\includegraphics[width=\textwidth]{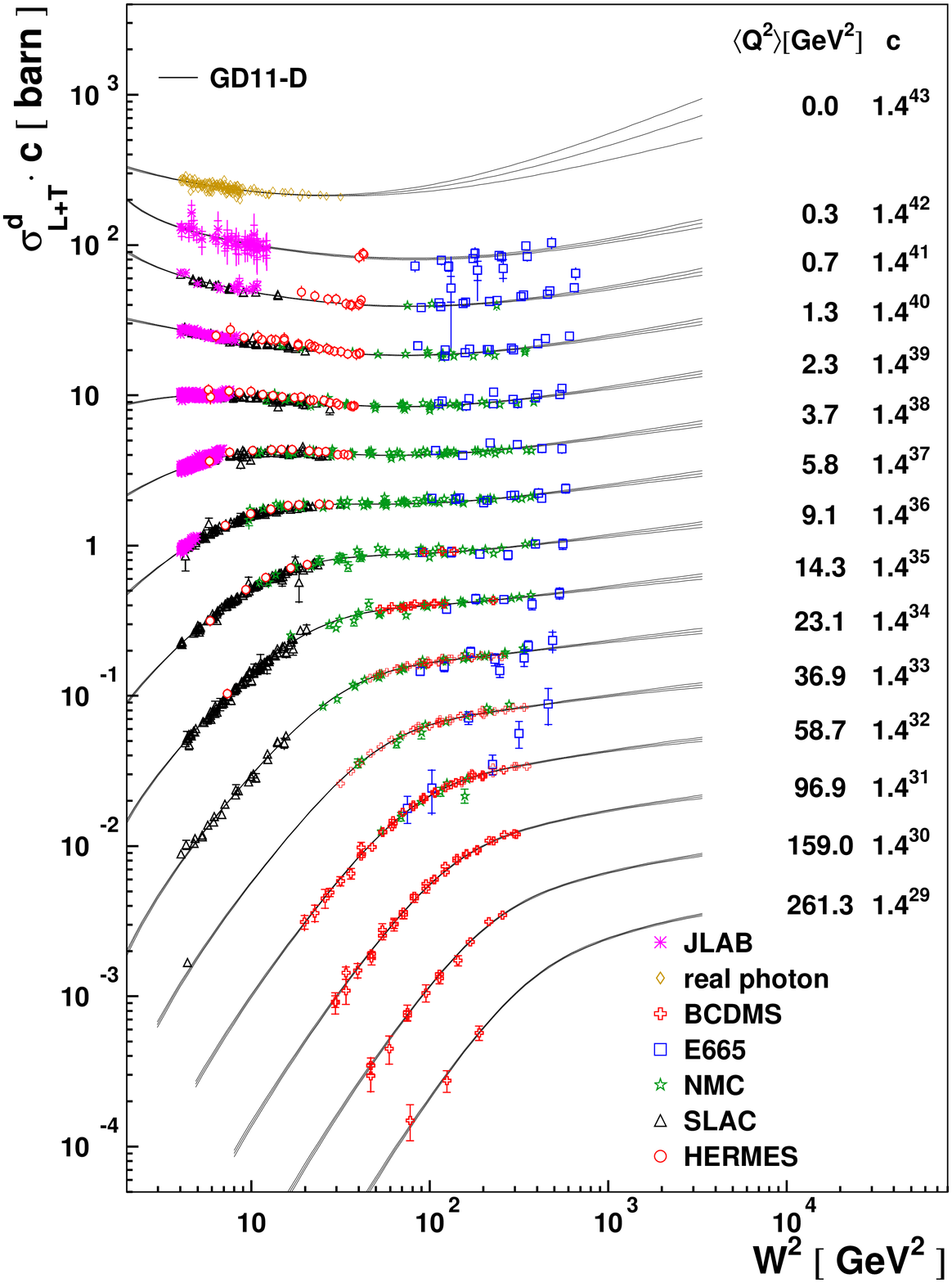}
\caption{\label{fig:sigltd0}
HERMES data for 
the photon-deuteron 
cross section $\sigma_{L+T}^{d}$  
  as a function of $W^2$, together with world data and the results from
  the GD11-D fit (central curves) and its uncertainties (outer curves), 
in bins of $Q^2$.
  The data points denoted 'real photon' are for
  photoproduction. 
Inner error bars are statistical uncertainties, while
outer error bars are total uncertainties
calculated as the sum in quadrature of all statistical and 
systematic uncertainties including normalization.
}
}


\subsection{\label{sec:Vasymmetry} Cross-section ratio $\sigma^d/\sigma^p$}


From the measured cross-section ratio, $\sigma^d/\sigma^p$,  
the ratio of the deuteron and proton 
structure functions, $F_2^d/F_2^p$, can be extracted. In the DIS regime 
$F_2^d/F_2^p$ is related to the ratio of the down and up 
quark distribution functions and imposes strong constraints 
on the $x$ dependence of the flavor 
composition of the nucleon. From the combination of $F_2^p$ and 
$F_2^d/F_2^p$ the non-singlet structure function $F_2^p - F_2^n$ 
and the Gottfried sum $S_G = \int (F_2^p - F_2^n) dx/x$ can be 
determined. For the latter, it was shown that the simple quark-model 
expectation of 1/3 is not reached~\cite{gottfriednmc1, gottfriednmc2} 
and hence the light quark sea is not flavor-symmetric.
We do not elaborate on these aspects here, as our data add only limited 
additional information in the DIS region. 

The cross-section ratio $\sigma^d/\sigma^p$ is determined year-by-year and 
then averaged. 
Thereby the normalization uncertainty is reduced and 
the effects of PID efficiencies and contaminations cancel. 
Because the deuteron and the proton data were partly taken with different 
beam conditions and resulting possible differences in absolute normalization, 
the ratio of the $F_2$-values given in Tables~\ref{tab:f2p} and \ref{tab:f2d}  
may differ slightly from the ratio values presented here.
The results for $\sigma^d/\sigma^p$ are listed in Tab.~\ref{tab:f2ratio}. 
The values  are shown as a function of $Q^2$ in bins of $x$ 
in Fig.~\ref{fig:sigdsigpx}  together with world data. 

\FIGURE{
\includegraphics[width=\textwidth]{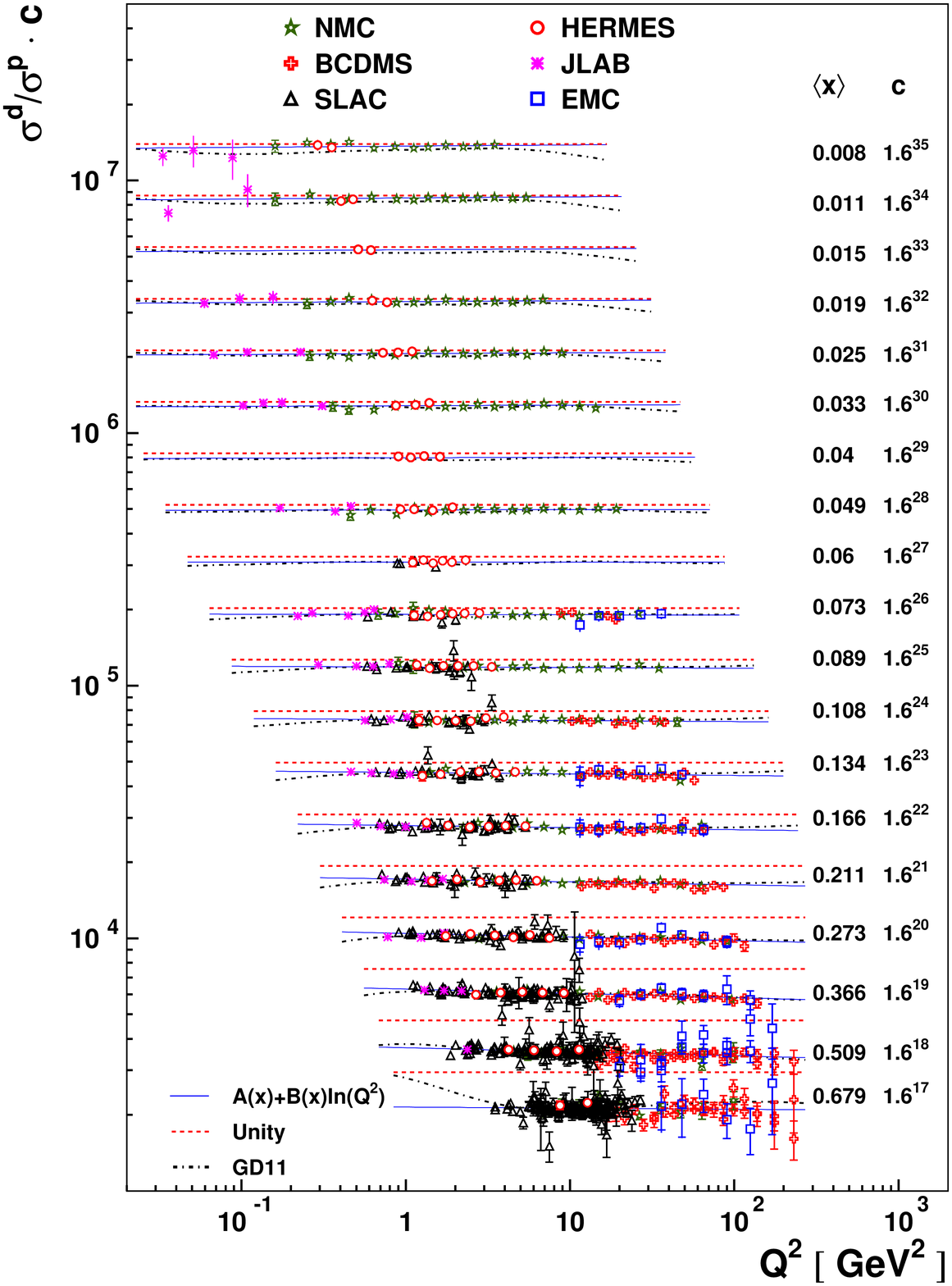}
\caption{\label{fig:sigdsigpx} 
The $Q^2$ dependence of $\sigma^d/\sigma^p$ for 
data from SLAC, HERMES, EMC, NMC, JLAB and BCDMS in bins of $\langle x\rangle$. 
The error bars on the data points represent  total uncertainties.
The data points are overlaid with 
the results from the fit described in section~\protect{\ref{sec:Vasymmetry}} 
(blue solid line), with the unity line (red dashed) 
and with the ratio of the GD11-D over GD11-P fits (green dash-dotted line). 
}}


\subsection{\label{app:fits}
Fits to world data to the cross section $\sigma_{L+T}^{p,d}$.}


Fits of the inclusive inelastic scattering 
cross sections on the proton and deuteron are performed.
The ALLM functional form selected for the fits is described in
Refs.~\cite{allm:1991} and \cite{allm:1997}. 
It is a 23-parameter Regge-motivated model of
$\sigma_{L+T}^{p,d}=\sigma_{L}^{p,d}+\sigma_{T}^{p,d}$,
valid for $W^2>4\,$GeV$^2$, {\it i.e.}, above 
the resonance region, and any $Q^2$ including the real photon point
($Q^2=0$).

These new fits to the photon-nucleon cross sections
$\sigma_{L+T}$ on proton (GD11-P) and deuteron (GD11-D) reflect the
recent world knowlege on the cross sections.
In order to present a self-consistent set, the values of $F_2$ 
to be used in the fit are
calculated from the measured cross sections using for all data the
parametrization $R_{1998}$. 
In cases where the measured cross sections were not given, 
the latter are reconstructed from the published values 
for $F_2$ using the values of $R$ used in calculating these $F_2$.
Both fits GD11-P and GD11-D are performed in the same manner.

The fit to proton data includes 2441  data points
from the SLAC  experiments E49a, E49b, E61, E87, E89a, 
E89b~\cite{whitlow:thesis}, NMC~\cite{nmc9610231},
the  combined HERA data~\cite{:2009wt},
 E665~\cite{e665},
 BCDMS~\cite{bcdms}, 
JLAB~\cite{Osipenko:2003ua,Malace:2009kw,Tvaskis:2010as},
the  HERMES data presented here, and real photon data~\cite{pdg2010}.
In the case of HERA, only data points with $Q^2<1000$~GeV$^2$ were used 
in order to 
minimize contributions from weak processes  to the cross section.
The fit to deuteron data includes 2497  data points from  
 the SLAC  experiments E49a, E49b, E61, E87, E89a, E89b, E139, 
E140~\cite{whitlow:thesis}, 
 NMC~\cite{nmc9610231}; 
 E665~\cite{e665}; 
 BCDMS~\cite{bcdms}; 
 JLAB~\cite{Osipenko:2005gt,Malace:2009kw,Tvaskis:2010as};
the  HERMES data  presented in this paper, and real photon data~\cite{pdg2010}. 
Tab.~\ref{tab:gd11data} lists in more detail the data sets, together with their
 $\langle x\rangle$ and $\langle Q^2\rangle$ ranges.

\TABLE{
\caption{\label{tab:gd11data}Data sets used in the GD11 fits.}
\begin{tabular}{|l|c|c|c|c|}
\hline
\hline
Experiment                               & \# of  & $\langle x\rangle$ range  & $\langle Q^2\rangle$ range &target\\
                                                   &       points       &   & [GeV$^2$] & \\
\hline
HERA (positron beam)        \cite{:2009wt}         & 456      & 0.621$\times 10^{-6}$ - 0.65 & 0.045 - 1000   & p\\
HERA (electron beam)        \cite{:2009wt}         & 95      & 0.13$\times 10^{-2}$ - 0.65  & 90     - 1000 & p\\
E665                        \cite{e665}            & 91       & 0.89$\times 10^{-3}$ - 0.39  & 0.23 - 64      & p, d  \\
NMC - 90 GeV                 \cite{nmc9610231}      & 73       & 0.78$\times 10^{-2}$ - 0.465 & 0.8  - 8.75    & p, d\\
NMC - 120 GeV                \cite{nmc9610231}     & 65       & 0.87$\times 10^{-2}$ - 0.475 & 1.18 - 14.4    & p, d\\
NMC - 200 GeV                \cite{nmc9610231}     & 75       & 0.35$\times 10^{-2}$ - 0.477 & 0.83 - 34.8    & p, d\\
NMC - 280 GeV                \cite{nmc9610231}     & 79       & 0.37$\times 10^{-2}$ - 0.479 & 1.27 - 62.3    & p, d\\
BCDMS - 100 GeV             \cite{bcdms}           & 58       & 0.07 -  0.75                & 7.5   - 75     & p \\
BCDMS - 120 GeV             \cite{bcdms}           & 62       & 0.07 -  0.75                & 8.75  - 99     & p, d\\
BCDMS - 200 GeV            \cite{bcdms}           & 57       & 0.07 - 0.75                 & 8.75  - 137    & p \\
BCDMS - 200 GeV             \cite{bcdms}           & 56       & 0.07 - 0.75                 & 17  - 137.5    & d \\
BCDMS - 280 GeV             \cite{bcdms}           & 52       & 0.1 - 0.75                  & 32.5  - 230    & p, d\\
SLAC E49a                   \cite{whitlow:thesis}  & 98       &  0.067 - 0.71               & 0.59 - 7.9     & p, d\\
SLAC E49b                   \cite{whitlow:thesis}  & 187      &  0.08 - 0.854               & 0.663 - 20     & p \\
SLAC E49b                   \cite{whitlow:thesis}  & 174      &  0.08 - 0.854               & 0.663 - 20     & d \\
SLAC E61                    \cite{whitlow:thesis}  & 25       &  0.065 - 0.326              & 0.58 - 1.68    & p \\
SLAC E61                    \cite{whitlow:thesis}  & 24       &  0.065 - 0.326              & 0.58 - 1.68    & d \\
SLAC E87                    \cite{whitlow:thesis}  & 94       & 0.313 - 0.832               & 3.96 - 19.7    & p, d\\
SLAC E89a                   \cite{whitlow:thesis}  & 72       &  0.31 - 0.9                 & 3.6 - 30.3     & p\\
SLAC E89a                   \cite{whitlow:thesis}  & 66       &  0.33 - 0.9                 & 3.86 - 30.2     & d\\
SLAC E89b                   \cite{whitlow:thesis}  & 98       & 0.063 - 0.809               & 0.887 - 18.4   & p\\
SLAC E89b                   \cite{whitlow:thesis}  & 81       & 0.063 - 0.807               & 0.887 - 18.4   & d\\
SLAC E139                   \cite{whitlow:thesis}  & 18       &  0.089 - 0.7               & 2 - 15 & d\\
SLAC E140                   \cite{whitlow:thesis}  & 38       & 0.2 - 0.5   & 1 - 10 & d \\
JLAB  E00-116              \cite{Malace:2009kw}   & 50       & 0.483 - 0.641               & 3.585 - 5.67   &p, d \\
JLAB CLAS                \cite{Osipenko:2003ua}  & 272      & 0.23 - 0.509                & 1.325 - 3.275  &p \\
JLAB CLAS                \cite{Osipenko:2005gt}  & 1018      & 0.122 - 0.617                & 0.475 - 5.125 &d \\
JLAB HALL C (Rosenbluth) \cite{Tvaskis:2010as}   & 5        & 0.25 - 0.8                  & 0.15 - 1.045   & p, d\\
JLAB HALL C (Model Dep.) \cite{Tvaskis:2010as}   & 50       & 0.9$\times 10^{-2}$  - 0.25  & 0.034 - 1.761  & p,  d\\
HERMES                      (this analysis)       & 81       & 0.69$\times 10^{-2}$ - 0.664 & 0.291 -12.78   & p, d\\
real photon                 \cite{pdg2010}         & 196      & -                            & -             & p\\
real photon                 \cite{pdg2010}         & 174      & -                            & -             & d\\
\hline
\hline
\end{tabular}
}

The fits are based on the minimization of the value of $\chi^2$ defined as
\begin{eqnarray}
\label{Eq:chi2}
\chi^2({\bf p},{\mbox{\boldmath $\nu$}}) 
 &=&\sum_{i,k}{\frac{[D_{i,k}(W^2,Q^2)\cdot(1+\delta_k{\mathbf \nu_k})
    -T({\bf{p}},W^2,Q^2)]^2 
  }{({\sigma_{i,k}^{stat}}^2+{\sigma_{i,k}^{syst}}^2)
\cdot(1+\delta_k\nu_k)^2}} +\sum_k{\nu_k^2} 
\nn
\\
&\approx&\sum_{i,k}{\frac{ [ D_{i,k}(W^2,Q^2) - T({\bf{p}},W^2,Q^2)  \cdot
     (1-\delta_k\nu_k) 
     ]^2 } 
{{\sigma_{i,k}^{stat}}^2+{\sigma_{i,k}^{syst}}^2}}+\sum_k{\nu_k^2} ~,
\end{eqnarray}
where  $D_{i,k}\pm \sigma_{i,k}^{stat}\pm\sigma_{i,k}^{syst}$ are
the values of $\sigma_{L+T}$ for 
data point $i$ within the data set $k$, 
$\delta_k$ is the normalization uncertainty in data set $k$ quoted by the
experiment, $\nu_k$ is a normalization parameter, $T({\bf p},W^2,Q^2)$ is the 
23-parameter ALLM functional form, ${\bf p}$ is the vector of functional 
parameters and {\boldmath$\nu$} is the vector of normalization parameters. 
The definition of $\chi^2$ takes into account point-by-point
statistical and systematic 
uncertainties. 
We agree that ideally one should treat the statistical and systematic
uncertainties separately, as the latter in most cases are to some degree 
correlated. However, almost all published data sets just give statistical, 
total, and normalization uncertainties. For that reason we have used in 
the fit the total uncertainties per point, plus the overall normalization.
Clearly the value of $\chi^2$ is influenced by this, but in practice the
influence on the resultant parameters seems minor, 
see for example Ref.~\cite{Martin:2001es}.
Additionally, for each data set $k$ a  normalization parameter $\nu_k$ is 
used to describe the normalization of all data points belonging to 
the data set.
Since the normalization parameters are assumed to normalize data
within the known normalization uncertainties, they are scaled with
these uncertainties and taken into account by a penalty term $\sum_k{\nu_k^2}$ to
control their variation 
according  to their standard deviations.

At each iteration of the $\chi^2$-minimization, the normalization
parameters $\nu_k$ are analytically determined:
\begin{equation}
\label{Eq:norm}
\nu_k
=\frac{\sum_i{\delta_kT_{i,k}(T_{i,k}-D_{i,k})/\sigma_{i,k}^2}}{
\sum_iT_{i,k}^2\delta_k^2/\sigma_{i,k}^2+1},
\end{equation}
where $\sigma_{i,k}^2={\sigma_{i,k}^{stat}}^2+{\sigma_{i,k}^{syst}}^2$.  
This equation is obtained by requiring $\partial \chi^2/\partial \nu_k=0$ 
in the context of the approximation for $\chi^2$ in 
Eq.~(\ref{Eq:chi2}). 
The analytical determination of the parameters $\nu_k$ reduces the 
number of free parameters in the $\chi^2$-minimization. 
The separate extraction of all normalization parameters is 
possible only because the parameters $\nu_k$ are 
uncorrelated.
At each step Eq.~(\ref{Eq:norm}) 
is substituted into the $\chi^2$-function, Eq.~(\ref{Eq:chi2}), and thus the
final fit  result is obtained by minimizing $\chi^2$ only with
respect to the functional parameters. Similar methods 
are used by others in order to save computing time~\cite{wallny,Stump:2001gu}.

The fits are performed using the \texttt{CERNLIB} package
\texttt{MINUIT} \cite{minuit} and the \texttt{MINUIT} 
extension~\cite{pumplin}. 
The resulting fit parameters are listed in Tab.~\ref{tab:parameters} together 
with the parameters obtained in the previous fits ALLM97~\cite{allm:1997} and 
GD07-P~\cite{gd07} 
 performed on proton data.
The  difference between GD07-P and GD11-P is the presence of
  HERMES, JLAB and the combined HERA data in the most recent fit. 
In Ref.~\cite{gd07} the
  uncertainties were calculated with an UP value~\cite{minuit} equal to
  24.7. 
That identical fit was redone here with the same value of UP=1
as the new fits in order to provide an appropriate comparison of 
its uncertainties.
The ALLM97 fit provided no uncertainties.
The uncertainties given in the table for GD07-P, GD11-P, and GD11-D 
correspond to the diagonal elements of the full covariance matrix,
which must be used to calculate uncertainties in $F_2$ or cross
sections. 
The $\chi^2/$ per degree of freedom for the GD11-P is 0.89 (it is 0.92 
for GD07-P),  
while for the deuteron fit  GD11-D it is 0.68.
The fitted relative normalization of HERMES proton (deuteron) data is +1.5\% 
(-2.2\%), 
well within the quoted uncertainty of 7.5\% (7.6\%). 
Tab.~\ref{tab:gd10normalizations} shows the fitted  relative normalizations 
for all 
the data sets used in the fit.

\subsection{\label{app:fitsf2ratio}
Fits to world data for the cross-section ratio $\sigma^d/\sigma^p$.}

The world data on the cross-section ratio $\sigma^d/\sigma^p$ is fitted using 
the functional form 
\begin{equation}
v\sigma^d/\sigma^p(x,Q^2)=A(x)+B(x)\ln {Q^2}, 
\end{equation}
with
\begin{eqnarray}
 A(x)&=&p_1+p_2x+p_3x^2+p_4x^3+p_5x^4,\nn\\
B(x)&=&p_6+p_7x+p_8x^2~,
\end{eqnarray}
 previously used by NMC in Ref.~\cite{nmcratiofit}.
It  includes eight free parameters $p_1,...,p_8$. 
The full data set consists of 
260 data points from NMC~\cite{Arneodo:1996kd}, 
621 data points from SLAC~\cite{whitlow:thesis}, 
49 data points from JLAB~\cite{Tvaskis:2010as}, 
81 data points from this analysis, 
159 data points from BCDMS, 
and 53 data points from EMC. 
The latter two data sets are taken from Ref.~\cite{whitlow:thesis}. 
The relative normalizations are also fitted using the same method as described 
in the 
previous section. The final results are listed in Tab.~\ref{tab:csratio}.
\TABLE{
{
\caption{\label{tab:csratio} 
Results of fit to the cross-section ratio $\sigma^d/\sigma^p$. 
The meaning of the parameters $p_1,...p_8$ and relative normalizations  is 
explained in the text.
}
}
\begin{tabular}{|c|c|c|}
\hline
\hline
parameter   &  value & uncertainty \\
\hline
    $p_1$     &    0.986  &     0.0021\\
    $p_2$     &   -0.684   &    0.034\\
    $p_3$     &    1.54    &   0.18    \\
    $p_4$     &   -2.93    &   0.36    \\
    $p_5$     &    1.96    &   0.22    \\
    $p_6$     &    0.0053 &   0.0011\\
    $p_7$     &    -0.0974 &   0.0087\\
    $p_8$     &   0.125       & 0.014       \\
\hline
normalization & value &  \\
\hline
    HERMES &    0.996    &    - \\
    NMC    &    0.999    &   -   \\
    BCDMS  &     1.010    &   -   \\
    SLAC  &     1.003    &    - \\
    JLAB   &     1.000    &    - \\
    EMC    &     0.995    &    -   \\
\hline
\hline
\end{tabular}                                              
}
The relative normalizations were all found to be well within the normalization 
uncertainties
 quoted by the experiments. 
In the case of HERMES data, it is found to be 0.996.
Fig.~\ref{fig:sigdsigpx} shows the world data of $\sigma^d/\sigma^p$ overlapped 
 with the fits described in this section and with the ratio 
of $F_2^d$ to $F_2^p$ derived from the GD11 fits.
While the ratio deviates from unity at large $x$, both GD11 and NMC-like fits 
describe the data well but start to deviate from each other outside the 
range where the ratio has been determined directly.


\section{\label{sec:summary}Summary}


This high-statistics measurement of the inelastic scattering cross section
on the proton and the deuteron by HERMES  provides data in the ranges 
$0.006\leq x\leq 0.9$ and 
0.1~GeV$^2\leq Q^2\leq$ 20 GeV$^2$ and contributes data in a kinematic region 
previously not covered, 
$Q^2\lesssim 1$ GeV$^2$ and $15$\,GeV$^2<W^2<45$\,GeV$^2$, 
corresponding to $0.006<x<0.04$.
In the region of overlap, HERMES data agree with the world data. 
The new results are fitted 
in conjunction with the other data, including those at the
photon point, to give a new parameterization of the structure functions
$F_2^p$ ($F_2^d$) for $Q^2=0$ to 30000~GeV$^2$ (230~GeV$^2$) and $x$-values 
$0.6\times 10^{-6}$ (0.89$\times 10^{-3}$) to 0.9.
The fitted relative normalizations of HERMES data are 
well within the quoted normalization uncertainties.

The proton and deuteron cross-section ratio was also determined.
This quantity has a much smaller point-to-point uncertainty because several
systematic uncertainties cancel in the ratio and its normalization uncertainty
is also much smaller.
The HERMES data show  excellent agreement with the other data and the fitted
normalization is  well within the quoted normalization uncertainty.
The cross-section ratio from HERMES is also fitted in conjunction 
with existing data from other experiments. 
The $Q^2$ dependence can be well described by a simple ($x$-dependent) $\ln Q^2$ 
term.


\begin{acknowledgments}

We gratefully acknowledge the DESY management for its support and the staff
at DESY and the collaborating institutions for their significant effort.
This work was supported by 
the Ministry of Economy and the Ministry of Education and Science of Armenia;
the FWO-Flanders and IWT, Belgium;
the Natural Sciences and Engineering Research Council of Canada;
the National Natural Science Foundation of China;
the Alexander von Humboldt Stiftung,
the German Bundesministerium f\"ur Bildung und Forschung (BMBF), and
the Deutsche Forschungsgemeinschaft (DFG);
the Italian Istituto Nazionale di Fisica Nucleare (INFN);
the MEXT, JSPS, and G-COE of Japan;
the Dutch Foundation for Fundamenteel Onderzoek der Materie (FOM);
the Russian Academy of Science and the Russian Federal Agency for 
Science and Innovations;
the U.K.~Engineering and Physical Sciences Research Council, 
the Science and Technology Facilities Council,
and the Scottish Universities Physics Alliance;
the U.S.~Department of Energy (DOE) and the National Science Foundation (NSF);
the Basque Foundation for Science (IKERBASQUE);
and the European Community Research Infrastructure Integrating Activity
under the FP7 "Study of Strongly Interacting Matter (HadronPhysics2, Grant
Agreement number 227431)".

\end{acknowledgments}


\begin{appendix}


\clearpage

\bibliography{f2paper}

\providecommand{\href}[2]{#2}\begingroup\raggedright\begin{thebibliography}{10}

\bibitem{sarkar2004}
R.~Devenish and A.~Cooper-Sarkar, {\it Deep inelastic scattering},  {\em Oxford
  University Press} (2004).

\bibitem{Badelek:1994fe}
B.~Badelek and J.~Kwiecinski, {\it {The low-$Q^2$, low-x region in
  electroproduction}},  {\em Rev. Mod. Phys.} {\bf 68} (1996) 445--471,
  [\href{http://xxx.lanl.gov/abs/hep-ph/9408318}{{\tt hep-ph/9408318}}].

\bibitem{allm:1991}
H.~Abramowicz, E.~M. Levin, A.~Levy, and U.~Maor, {\it {A parametrization of
  $\sigma_T$ ($\gamma^*$ p) above the resonance region for $Q^2 \geq 0$}},
  {\em Phys. Lett.} {\bf B269} (1991) 465--476.

\bibitem{Aid:1996au}
{\bf H1} Collaboration, S.~Aid {\em et~al.}, {\it {A measurement and QCD
  analysis of the proton structure function $F_2(x,Q^2)$ at HERA}},  {\em Nucl.
  Phys.} {\bf B470} (1996) 3--40,
  [\href{http://xxx.lanl.gov/abs/hep-ex/9603004}{{\tt hep-ex/9603004}}].

\bibitem{Adloff:1997mf}
{\bf H1} Collaboration, C.~Adloff {\em et~al.}, {\it {A measurement of the
  proton structure function $F_2(x,Q^2)$ at low x and low $Q^2$ at HERA}},
  {\em Nucl. Phys.} {\bf B497} (1997) 3--30,
  [\href{http://xxx.lanl.gov/abs/hep-ex/9703012}{{\tt hep-ex/9703012}}].

\bibitem{Adloff:1999ah}
{\bf H1} Collaboration, C.~Adloff {\em et~al.}, {\it {Measurement of neutral
  and charged current cross-sections in positron-proton collisions at large
  momentum transfer}},  {\em Eur. Phys. J.} {\bf C13} (2000) 609--639,
  [\href{http://xxx.lanl.gov/abs/hep-ex/9908059}{{\tt hep-ex/9908059}}].

\bibitem{Adloff:2000qj}
{\bf H1} Collaboration, C.~Adloff {\em et~al.}, {\it {Measurement of neutral
  and charged current cross sections in electron - proton collisions at high
  $Q^{2}$}},  {\em Eur. Phys. J.} {\bf C19} (2001) 269--288,
  [\href{http://xxx.lanl.gov/abs/hep-ex/0012052}{{\tt hep-ex/0012052}}].

\bibitem{Adloff:2003uh}
{\bf H1} Collaboration, C.~Adloff {\em et~al.}, {\it {Measurement and QCD
  analysis of neutral and charged current cross sections at HERA}},  {\em Eur.
  Phys. J.} {\bf C30} (2003) 1--32,
  [\href{http://xxx.lanl.gov/abs/hep-ex/0304003}{{\tt hep-ex/0304003}}].

\bibitem{zeuslowx}
{\bf ZEUS} Collaboration, M.~Derrick {\em et~al.}, {\it {Measurement of the
  proton structure function $F_2$ in e p scattering at HERA}},  {\em Phys.
  Lett.} {\bf B316} (1993) 412--426.

\bibitem{Derrick:1995ef}
{\bf ZEUS} Collaboration, M.~Derrick {\em et~al.}, {\it {Measurement of the
  proton structure function ${F_2}$ at low ${x}$ and low ${Q^2}$ at HERA}},
  {\em Z. Phys.} {\bf C69} (1996) 607--620,
  [\href{http://xxx.lanl.gov/abs/hep-ex/9510009}{{\tt hep-ex/9510009}}].

\bibitem{Derrick:1996hn}
{\bf ZEUS} Collaboration, M.~Derrick {\em et~al.}, {\it {Measurement of the
  $F_2$ structure function in deep inelastic e$^+$ p scattering using 1994 data
  from the ZEUS detector at HERA}},  {\em Z. Phys.} {\bf C72} (1996) 399--424,
  [\href{http://xxx.lanl.gov/abs/hep-ex/9607002}{{\tt hep-ex/9607002}}].

\bibitem{Breitweg:1997hz}
{\bf ZEUS} Collaboration, J.~Breitweg {\em et~al.}, {\it {Measurement of the
  proton structure function $F_2$ and $\sigma$(tot)($\gamma^*$ p) at low $Q^2$
  and very low x at HERA}},  {\em Phys. Lett.} {\bf B407} (1997) 432--448,
  [\href{http://xxx.lanl.gov/abs/hep-ex/9707025}{{\tt hep-ex/9707025}}].

\bibitem{Breitweg:2000yn}
{\bf ZEUS} Collaboration, J.~Breitweg {\em et~al.}, {\it {Measurement of the
  proton structure function $F_2$ at very low $Q^2$ at HERA}},  {\em Phys.
  Lett.} {\bf B487} (2000) 53--73,
  [\href{http://xxx.lanl.gov/abs/hep-ex/0005018}{{\tt hep-ex/0005018}}].

\bibitem{Chekanov:2001qu}
{\bf ZEUS} Collaboration, S.~Chekanov {\em et~al.}, {\it {Measurement of the
  neutral current cross section and $F_2$ structure function for deep inelastic
  e$^+$ p scattering at HERA}},  {\em Eur. Phys. J.} {\bf C21} (2001) 443--471,
  [\href{http://xxx.lanl.gov/abs/hep-ex/0105090}{{\tt hep-ex/0105090}}].

\bibitem{bcdms}
{\bf BCDMS} Collaboration, A.~C. Benvenuti {\em et~al.}, {\it {A HIGH
  STATISTICS MEASUREMENT OF THE PROTON STRUCTURE FUNCTIONS $F_2(x, Q^2$) AND R
  FROM DEEP INELASTIC MUON SCATTERING AT HIGH $Q^2$}},  {\em Phys. Lett.} {\bf
  B223} (1989) 485--489.

\bibitem{emceffect}
{\bf European Muon} Collaboration, J.~J. Aubert {\em et~al.}, {\it
  {Measurements of the Nucleon Structure Functions $F(2)N$ in Deep Inelastic
  Muon Scattering from Deuterium and Comparison With Those from Hydrogen and
  Iron}},  {\em Nucl. Phys.} {\bf B293} (1987) 740--786.

\bibitem{nmc9610231}
{\bf New Muon} Collaboration, M.~Arneodo {\em et~al.}, {\it {Measurement of the
  proton and deuteron structure functions, $F_2^p$ and $F_2^d$, and of the
  ratio $\sigma_L/\sigma_T$}},  {\em Nucl. Phys.} {\bf B483} (1997) 3--43,
  [\href{http://xxx.lanl.gov/abs/hep-ph/9610231}{{\tt hep-ph/9610231}}].

\bibitem{e665}
{\bf E665} Collaboration, M.~R. Adams {\em et~al.}, {\it {Proton and deuteron
  structure functions in muon scattering at 470 GeV}},  {\em Phys. Rev.} {\bf
  D54} (1996) 3006--3056.

\bibitem{whitlow:thesis}
L.~W. Whitlow, {\it Deep inelastic structure functions from electron scattering
  on hydrogen, deuterium, and iron at {0.6 GeV$^2\leq Q^2 \leq$ 30 GeV$^2$}},
  {\em SLAC-0357}.

\bibitem{Osipenko:2003ua}
M.~Osipenko {\em et~al.}, {\it {The proton structure function $F_2$ with
  CLAS}},  \href{http://xxx.lanl.gov/abs/[hep-ex/0309052]}{{\tt
  [hep-ex/0309052]}}.

\bibitem{Osipenko:2005gt}
{\bf CLAS} Collaboration, M.~Osipenko {\em et~al.}, {\it {Measurement of the
  deuteron structure function $F_2$ in the resonance region and evaluation of
  its moments}},  {\em Phys. Rev.} {\bf C73} (2006) 045205,
  [\href{http://xxx.lanl.gov/abs/hep-ex/0506004}{{\tt hep-ex/0506004}}].

\bibitem{Malace:2009kw}
{\bf Jefferson Lab E00-115} Collaboration, S.~P. Malace {\em et~al.}, {\it
  {Applications of quark-hadron duality in the $F_2$ structure function}},
  {\em Phys. Rev.} {\bf C80} (2009) 035207,
  [\href{http://xxx.lanl.gov/abs/nucl-ex/0905.2374}{{\tt nucl-ex/0905.2374}}].

\bibitem{Tvaskis:2010as}
V.~Tvaskis {\em et~al.}, {\it {The proton and deuteron $F_2$ structure function
  at low $Q^2$}},  {\em Phys. Rev.} {\bf C81} (2010) 055207,
  [\href{http://xxx.lanl.gov/abs/1002.1669}{{\tt arXiv:1002.1669}}].

\bibitem{Ackerstaff:1998av}
{\bf HERMES} Collaboration, K.~Ackerstaff {\em et~al.}, {\it {The HERMES
  Spectrometer}},  {\em Nucl. Instrum. Meth.} {\bf A417} (1998) 230--265,
  [\href{http://xxx.lanl.gov/abs/hep-ex/9806008}{{\tt hep-ex/9806008}}].

\bibitem{allm:1997}
H.~Abramowicz and A.~Levy, {\it {The ALLM Parameterization of
  $\sigma(tot)(\gamma^*$ p): An Update}},
  \href{http://xxx.lanl.gov/abs/[hep-ph/9712415]}{{\tt [hep-ph/9712415]}}.

\bibitem{close}
F.~Close, {\em An Introduction to Quarks and Partons}.
\newblock London Academic Press, 1979.

\bibitem{handconvention1}
L.~N. Hand, {\it {Experimental Investigation of Pion Electroproduction}},  {\em
  Phys. Rev.} {\bf 129} (1963) 1834--1846.

\bibitem{handconvention2}
H.~Abramowicz and A.~Caldwell, {\it {HERA collider physics}},  {\em Rev. Mod.
  Phys.} {\bf 71} (1999) 1275--1410,
  [\href{http://xxx.lanl.gov/abs/hep-ex/9903037}{{\tt hep-ex/9903037}}].

\bibitem{Akopov:2000qi}
N.~Akopov {\em et~al.}, {\it {The HERMES dual-radiator ring imaging Cerenkov
  detector}},  {\em Nucl. Instrum. Meth.} {\bf A479} (2002) 511--530,
  [\href{http://xxx.lanl.gov/abs/physics/0104033}{{\tt physics/0104033}}].

\bibitem{rich2}
E.~Aschenauer {\em et~al.}, {\it {Optical characterization of n = 1.03 silica
  aerogel used as radiator in the RICH of HERMES}},  {\em Nucl. Instrum. Meth.}
  {\bf A440} (2000) 338--347.

\bibitem{Benisch:2001rr}
T.~Benisch {\em et~al.}, {\it {The luminosity monitor of the HERMES experiment
  at DESY}},  {\em Nucl. Instrum. Meth.} {\bf A471} (2001) 314--324.

\bibitem{Abe:R1998}
{\bf E143} Collaboration, K.~Abe {\em et~al.}, {\it {Measurements of R =
  $\sigma_L/\sigma_T$ for 0.03 $< x < $ 0.1 and fit to world data}},  {\em
  Phys. Lett.} {\bf B452} (1999) 194--200,
  [\href{http://xxx.lanl.gov/abs/hep-ex/9808028}{{\tt hep-ex/9808028}}].

\bibitem{r1990}
L.~W. Whitlow, S.~Rock, A.~Bodek, E.~M. Riordan, and S.~Dasu, {\it {A precise
  extraction of $R = \sigma_L / \sigma_T$ from a global analysis of the SLAC
  deep inelastic e-p and e-d scattering cross sections}},  {\em Phys. Lett.}
  {\bf B250} (1990) 193--198.

\bibitem{nmcratiofit}
{\bf New Muon} Collaboration, P.~Amaudruz {\em et~al.}, {\it {The ratio $F_2^n
  / F_2^p$ in deep inelastic muon scattering}},  {\em Nucl. Phys.} {\bf B371}
  (1992) 3--31.

\bibitem{radgen}
{I. Akushevich, H. B{\"o}ttcher, and D. Ryckbosch}, {\it {RADGEN 1.0: Monte
  Carlo Generator for Radiative Events in DIS on Polarized and Unpolarized
  Targets}},  \href{http://xxx.lanl.gov/abs/[hep-ph/9906408]}{{\tt
  [hep-ph/9906408]}}.

\bibitem{formfactorp}
S.~I. Bilenkaya, S.~M. Bilenkii, Y.~M. Kazarinov, and L.~I. Lapidus, {\it {The
  Proton Electromagnetic Form Factor and Heavy Hypothetical Particles}},  {\em
  Pisma Zh. Eksp. Teor. Fiz.} {\bf 19} (1973) 613--616.

\bibitem{formfactorn}
G.~H{\"o}hler {\em et~al.}, {\it {ANALYSIS OF ELECTROMAGNETIC NUCLEON
  FORM-FACTORS}},  {\em Nucl. Phys.} {\bf B114} (1976) 505--534.

\bibitem{Arrington:2007ux}
J.~Arrington, W.~Melnitchouk, and J.~A. Tjon, {\it {Global analysis of proton
  elastic form factor data with two-photon exchange corrections}},  {\em Phys.
  Rev.} {\bf C76} (2007) 035205, [\href{http://xxx.lanl.gov/abs/0707.1861}{{\tt
  arXiv:0707.1861}}].

\bibitem{hermesg1}
{\bf HERMES} Collaboration, A.~Airapetian {\em et~al.}, {\it {Precise
  determination of the spin structure function $g_1$ of the proton, deuteron
  and neutron}},  {\em Phys. Rev.} {\bf D75} (2007) 012007,
  [\href{http://xxx.lanl.gov/abs/hep-ex/0609039}{{\tt hep-ex/0609039}}].

\bibitem{hermes:erratum}
{\bf HERMES} Collaboration, K.~Ackerstaff {\em et~al.}, {\it {Nuclear effects
  on R = $\sigma_L/\sigma_T$ in deep-inelastic scattering}},  {\em Phys. Lett.}
  {\bf B475} (2000) 386--394,
  [\href{http://xxx.lanl.gov/abs/hep-ex/9910071}{{\tt hep-ex/9910071}}].

\bibitem{HERMES:g1p}
{\bf HERMES} Collaboration, A.~Airapetian {\em et~al.}, {\it {Measurement of
  the proton spin structure function $g_1^p$ with a pure hydrogen target}},
  {\em Phys. Lett.} {\bf B442} (1998) 484--492,
  [\href{http://xxx.lanl.gov/abs/hep-ex/9807015}{{\tt hep-ex/9807015}}].

\bibitem{gottfriednmc1}
{\bf New Muon} Collaboration, P.~Amaudruz {\em et~al.}, {\it {Gottfried Sum
  from the Ratio $F_2^n / F_2^p$}},  {\em Phys. Rev. Lett.} {\bf 66} (1991)
  2712--2715.

\bibitem{gottfriednmc2}
{\bf New Muon} Collaboration, M.~Arneodo {\em et~al.}, {\it {Reevaluation of
  the Gottfried sum}},  {\em Phys. Rev.} {\bf D50} (1994) 1--3.

\bibitem{:2009wt}
{\bf H1 and ZEUS} Collaboration, F.~D. Aaron {\em et~al.}, {\it {Combined
  Measurement and QCD Analysis of the Inclusive e$^{\pm}$p Scattering Cross
  Sections at HERA}},  {\em JHEP} {\bf 01} (2010) 109,
  [\href{http://xxx.lanl.gov/abs/0911.0884}{{\tt arXiv:0911.0884}}].

\bibitem{pdg2010}
{\bf Particle Data Group} Collaboration, K.~Nakamura {\em et~al.}, {\it {Review
  of Particle Physics}},  {\em J. Phys.} {\bf G37} (2010) 075021.

\bibitem{Martin:2001es}
A.~D. Martin, R.~G. Roberts, W.~J. Stirling, and R.~S. Thorne, {\it {MRST2001:
  Partons and alpha(s) from precise deep inelastic scattering and Tevatron jet
  data}},  {\em Eur. Phys. J.} {\bf C23} (2002) 73--87,
  [\href{http://xxx.lanl.gov/abs/hep-ph/0110215}{{\tt hep-ph/0110215}}].

\bibitem{wallny}
R.~Wallny, {\it {A Measurement of the Gluon Distribution in the Proton and of
  the Strong Coupling Constant $\alpha_s$ from Inclusive Deep-Inelastic
  Scattering}},  {\em Universit{\"a}t Z{\"u}rich, Ph.D. thesis} (2001).

\bibitem{Stump:2001gu}
D.~Stump {\em et~al.}, {\it {Uncertainties of predictions from parton
  distribution functions. I. The Lagrange multiplier method}},  {\em Phys.
  Rev.} {\bf D65} (2001) 014012,
  [\href{http://xxx.lanl.gov/abs/hep-ph/0101051}{{\tt hep-ph/0101051}}].

\bibitem{minuit}
F.~James, {\it {MINUIT, Reference Manual, Version 94.1, CERN Program Library
  Long Writeup D506 (European Organization for Nuclear Research, Geneva,
  1994)}},  {\em MINUIT, CERN Program Library Long Writeup D506} (1994).

\bibitem{pumplin}
J.~Pumplin, D.~R. Stump, and W.~K. Tung, {\it {Multivariate fitting and the
  error matrix in global analysis of data}},  {\em Phys. Rev.} {\bf D65} (2001)
  014011, [\href{http://xxx.lanl.gov/abs/hep-ph/0008191}{{\tt
  hep-ph/0008191}}].

\bibitem{gd07}
D.~Gabbert and L.~De~Nardo, {\it {New Global Fit to the Total Photon-Proton
  Cross-Section $\sigma_{L+T}$ and to the Structure Function $F_2$}},
  \href{http://xxx.lanl.gov/abs/[hep-ph/0708.3196]}{{\tt [hep-ph/0708.3196]}}.

\bibitem{Arneodo:1996kd}
{\bf New Muon} Collaboration, M.~Arneodo {\em et~al.}, {\it {Accurate
  measurement of $F_2^d/F_2^p$ and $R^d-R^p$}},  {\em Nucl. Phys.} {\bf B487}
  (1997) 3--26, [\href{http://xxx.lanl.gov/abs/hep-ex/9611022}{{\tt
  hep-ex/9611022}}].

\end{thebibliography}\endgroup
\newpage
\section{Tables}

\TABLE{
{\renewcommand{\baselinestretch}{1.}
\caption{\label{tab:f2p} 
Results on the differential Born cross section $\frac{d^2\sigma^p}{dx\,dQ^2}$ 
and $F_2^p$. The statistical uncertainty $\delta_{stat.}$  and 
the systematic uncertainties 
$\delta_{PID}$ (particle identification), 
 $\delta_{model}$ (model dependence outside the acceptance),
 $\delta_{mis.}$ (misalignment), and
$\delta_{rad.}$ (Bethe-Heitler efficiencies)  
are given in percent. Corresponding $x$ bin numbers and $Q^2$ bin numbers 
and the average values $\langle x \rangle$ and $\langle {Q^2} \rangle$  are 
listed in the 
first four columns. 
The overall normalization uncertainty is 7.6\%. 
The structure function $F_2^p$  is derived using the parameterization 
$R=R_{1998}$~\cite{Abe:R1998}.
}
}
\begin{tabular}{|c|c|c|c|c|c|r|r|r|r|r|}
\hline
\hline
$x$ bin&$Q^2$ bin & $\langle x \rangle$ & $\langle Q^2 \rangle\,$  
&$\displaystyle\frac{d^2\sigma^p}{dx\,dQ^2}\,$ & $ F_2^p$& $\delta_{stat.}$ 
 & $\delta_{PID}$ &$\delta_{model}$ & $\delta_{mis.}$ & $\delta_{rad.}$\\
       &         &                       &  [GeV$^2$]              
& [nb/GeV$^2$] & & [\%] & [\%] & [\%] & [\%] & [\%]\\
\hline

1 & A & 0.0069 & 0.291 & 0.323$\times 10^5$ & 0.162 & 1.60 & 0.69 & 4.30 & 5.36 & 1.75\\
2 & A & 0.0102 & 0.404 & 0.143$\times 10^5$ & 0.200 & 1.35 & 0.87 & 2.79 & 3.95 & 0.48\\
3 & A & 0.0142 & 0.514 & 0.732$\times 10^4$ & 0.219 & 0.91 & 0.61 & 2.22 & 4.36 & 0.08\\
4 & A & 0.0186 & 0.626 & 0.432$\times 10^4$ & 0.241 & 0.99 & 0.36 & 1.85 & 3.89 & 0.10\\
5 & A & 0.0244 & 0.725 & 0.307$\times 10^4$ & 0.277 & 0.83 & 0.10 & 1.66 & 4.66 & 0.24\\
6 & A & 0.0325 & 0.872 & 0.182$\times 10^4$ & 0.299 & 0.88 & 0.01 & 1.54 & 3.78 & 0.39\\
7 & A & 0.0398 & 0.903 & 0.157$\times 10^4$ & 0.310 & 1.32 & 0.05 & 1.71 & 3.92 & 0.71\\
8 & A & 0.0486 & 0.929 & 0.141$\times 10^4$ & 0.334 & 2.15 & 0.07 & 2.76 & 6.43 & 0.27\\
9 & A & 0.0595 & 1.102 & 0.830$\times 10^3$ & 0.334 & 2.44 & 0.05 & 1.41 & 3.17 & 0.60\\
10 & A & 0.0724 & 1.127 & 0.764$\times 10^3$ & 0.368 & 2.39 & 0.06 & 1.20 & 5.49 & 0.71\\
11 & A & 0.0882 & 1.167 & 0.585$\times 10^3$ & 0.351 & 2.25 & 0.07 & 0.94 & 3.45 & 0.67\\
12 & A & 0.1078 & 1.205 & 0.498$\times 10^3$ & 0.373 & 3.23 & 0.07 & 0.85 & 5.93 & 0.54\\
13 & A & 0.1330 & 1.263 & 0.375$\times 10^3$ & 0.369 & 2.58 & 0.07 & 0.92 & 2.56 & 0.83\\
14 & A & 0.1657 & 1.337 & 0.287$\times 10^3$ & 0.383 & 2.25 & 0.06 & 1.15 & 7.52 & 1.27\\
15 & A & 0.2093 & 1.447 & 0.177$\times 10^3$ & 0.341 & 1.75 & 0.06 & 1.92 & 2.99 & 0.62\\
16 & A & 0.2642 & 1.749 & 0.997$\times 10^2$ & 0.352 & 1.10 & 0.05 & 2.87 & 3.29 & 0.55\\
17 & A & 0.3477 & 2.672 & 0.244$\times 10^2$ & 0.269 & 1.22 & 0.05 & 2.15 & 7.62 & 0.15\\
18 & A & 0.4605 & 4.224 & 0.475$\times 10$ & 0.179 & 1.09 & 0.04 & 1.83 & 4.10 & 0.08\\
19 & A & 0.6346 & 8.688 & 0.266 & 0.064 & 2.62 & 0.08 & 1.31 & 1.17 & 0.06\\
1 & B & 0.0085 & 0.354 & 0.205$\times 10^5$ & 0.191 & 1.45 & 0.93 & 3.31 & 3.75 & 1.35\\
2 & B & 0.0117 & 0.477 & 0.986$\times 10^4$ & 0.228 & 0.85 & 1.22 & 2.77 & 2.07 & 1.44\\
3 & B & 0.0152 & 0.612 & 0.492$\times 10^4$ & 0.243 & 0.68 & 1.50 & 2.53 & 1.03 & 0.76\\
4 & B & 0.0193 & 0.770 & 0.262$\times 10^4$ & 0.261 & 0.70 & 1.80 & 2.27 & 0.79 & 0.51\\
5 & B & 0.0250 & 0.896 & 0.169$\times 10^4$ & 0.272 & 0.56 & 1.07 & 1.83 & 0.56 & 0.41\\
6 & B & 0.0327 & 1.144 & 0.878$\times 10^3$ & 0.295 & 0.71 & 0.92 & 1.71 & 0.64 & 0.15\\
7 & B & 0.0401 & 1.074 & 0.103$\times 10^4$ & 0.315 & 1.09 & 0.01 & 1.46 & 2.61 & 0.15\\
8 & B & 0.0490 & 1.123 & 0.862$\times 10^3$ & 0.326 & 1.06 & 0.03 & 1.37 & 2.47 & 0.23\\
9 & B & 0.0596 & 1.281 & 0.591$\times 10^3$ & 0.343 & 2.00 & 0.06 & 0.92 & 2.26 & 0.56\\
10 & B & 0.0725 & 1.359 & 0.466$\times 10^3$ & 0.349 & 1.72 & 0.06 & 0.65 & 3.96 & 0.40\\
11 & B & 0.0884 & 1.399 & 0.407$\times 10^3$ & 0.371 & 1.78 & 0.06 & 0.59 & 3.58 & 0.14\\
\hline
\hline
\end{tabular}                                              
}

\addtocounter{table}{-1}                                   
\begingroup                                                
\TABLE{
{\caption{ -- continued.}}
\begin{tabular}{|c|c|c|c|c|c|r|r|r|r|r|}                   
\hline
\hline
$x$ bin&$Q^2$ bin & $\langle x \rangle$ & $\langle Q^2 \rangle\,$   
&$\displaystyle\frac{d^2\sigma^p}{dx\,dQ^2}$
 & $ F_2^p$& $\delta_{stat.}$& $\delta_{PID}$ & $\delta_{model}$ & $\delta_{mis.}$ 
& $\delta_{rad.}$ \\
       &         &                       &  [GeV$^2$]              & [nb/GeV$^2$] & 
& [\%] & [\%] & [\%] & [\%] & [\%]\\
\hline                                                     
12 & B & 0.1082 & 1.554 & 0.280$\times 10^3$ & 0.374 & 1.14 & 0.06 & 0.53 & 4.34 & 0.15\\
13 & B & 0.1333 & 1.632 & 0.215$\times 10^3$ & 0.372 & 1.67 & 0.06 & 0.56 & 1.84 & 0.35\\
14 & B & 0.1661 & 1.811 & 0.144$\times 10^3$ & 0.373 & 0.97 & 0.05 & 0.57 & 4.10 & 0.14\\
15 & B & 0.2104 & 2.050 & 0.857$\times 10^2$ & 0.352 & 1.38 & 0.05 & 0.64 & 2.80 & 0.26\\
16 & B & 0.2722 & 2.471 & 0.432$\times 10^2$ & 0.329 & 0.95 & 0.05 & 0.85 & 3.85 & 0.17\\
17 & B & 0.3620 & 3.790 & 0.110$\times 10^2$ & 0.268 & 0.81 & 0.04 & 0.64 & 2.21 & 0.06\\
18 & B & 0.4900 & 6.026 & 0.180$\times 10$ & 0.156 & 0.62 & 0.04 & 0.69 & 0.94 & 0.02\\
19 & B & 0.6641 & 12.778 & 0.795$\times 10^{-1}$ & 0.047 & 2.19 & 0.05 & 0.67 & 0.86 & 0.10\\
5 & C & 0.0270 & 1.092 & 0.102$\times 10^4$ & 0.289 & 0.76 & 2.39 & 2.20 & 1.06 & 0.18\\
6 & C & 0.0338 & 1.391 & 0.531$\times 10^3$ & 0.307 & 1.08 & 2.98 & 2.32 & 0.28 & 0.29\\
7 & C & 0.0401 & 1.293 & 0.629$\times 10^3$ & 0.313 & 0.83 & 0.38 & 1.54 & 2.71 & 0.17\\
8 & C & 0.0490 & 1.460 & 0.445$\times 10^3$ & 0.327 & 0.70 & 0.17 & 1.18 & 1.29 & 0.12\\
9 & C & 0.0596 & 1.466 & 0.419$\times 10^3$ & 0.339 & 1.10 & 0.01 & 0.80 & 1.73 & 0.45\\
10 & C & 0.0726 & 1.627 & 0.305$\times 10^3$ & 0.353 & 1.08 & 0.04 & 0.56 & 2.05 & 0.25\\
11 & C & 0.0885 & 1.691 & 0.247$\times 10^3$ & 0.351 & 1.09 & 0.05 & 0.49 & 1.43 & 0.06\\
12 & C & 0.1082 & 2.013 & 0.151$\times 10^3$ & 0.367 & 0.95 & 0.05 & 0.41 & 2.70 & 0.06\\
13 & C & 0.1334 & 2.158 & 0.109$\times 10^3$ & 0.359 & 1.04 & 0.05 & 0.39 & 2.17 & 0.15\\
14 & C & 0.1662 & 2.458 & 0.703$\times 10^2$ & 0.363 & 1.15 & 0.05 & 0.35 & 0.98 & 0.12\\
15 & C & 0.2105 & 2.851 & 0.415$\times 10^2$ & 0.355 & 0.73 & 0.04 & 0.33 & 2.40 & 0.06\\
16 & C & 0.2724 & 3.479 & 0.195$\times 10^2$ & 0.316 & 0.66 & 0.04 & 0.31 & 1.74 & 0.09\\
17 & C & 0.3622 & 5.125 & 0.520$\times 10$ & 0.250 & 0.60 & 0.04 & 0.23 & 1.84 & 0.02\\
18 & C & 0.4957 & 8.280 & 0.821 & 0.147 & 0.66 & 0.05 & 0.23 & 2.23 & 0.00\\
7 & D & 0.0408 & 1.615 & 0.349$\times 10^3$ & 0.317 & 0.72 & 2.58 & 2.08 & 0.53 & 0.35\\
8 & D & 0.0496 & 1.927 & 0.210$\times 10^3$ & 0.323 & 0.65 & 2.49 & 1.74 & 0.68 & 0.20\\
9 & D & 0.0596 & 1.675 & 0.297$\times 10^3$ & 0.337 & 1.22 & 0.04 & 0.79 & 1.88 & 0.18\\
10 & D & 0.0726 & 1.928 & 0.199$\times 10^3$ & 0.352 & 1.42 & 0.01 & 0.55 & 2.13 & 0.16\\
11 & D & 0.0885 & 2.080 & 0.150$\times 10^3$ & 0.354 & 0.91 & 0.03 & 0.47 & 1.21 & 0.16\\
12 & D & 0.1083 & 2.490 & 0.893$\times 10^2$ & 0.364 & 0.99 & 0.04 & 0.37 & 1.50 & 0.13\\
13 & D & 0.1334 & 2.795 & 0.614$\times 10^2$ & 0.372 & 0.82 & 0.04 & 0.31 & 2.67 & 0.10\\
14 & D & 0.1663 & 3.195 & 0.377$\times 10^2$ & 0.358 & 0.88 & 0.04 & 0.27 & 1.91 & 0.08\\
15 & D & 0.2106 & 3.709 & 0.219$\times 10^2$ & 0.343 & 0.68 & 0.05 & 0.22 & 1.80 & 0.09\\
16 & D & 0.2726 & 4.527 & 0.106$\times 10^2$ & 0.315 & 0.84 & 0.05 & 0.19 & 2.14 & 0.02\\
17 & D & 0.3624 & 6.797 & 0.263$\times 10$ & 0.242 & 0.57 & 0.04 & 0.13 & 1.63 & 0.01\\
18 & D & 0.5038 & 11.344 & 0.350 & 0.132 & 0.98 & 0.07 & 0.32 & 1.61 & 0.01\\
9 & E & 0.0597 & 1.906 & 0.210$\times 10^3$ & 0.333 & 0.93 & 0.30 & 0.87 & 0.78 & 0.17\\
10 & E & 0.0726 & 2.276 & 0.128$\times 10^3$ & 0.346 & 0.93 & 0.22 & 0.62 & 1.53 & 0.16\\
11 & E & 0.0885 & 2.587 & 0.863$\times 10^2$ & 0.351 & 0.70 & 0.09 & 0.47 & 1.33 & 0.03\\
12 & E & 0.1083 & 3.071 & 0.519$\times 10^2$ & 0.355 & 0.58 & 0.03 & 0.37 & 1.49 & 0.01\\
\hline
\hline
\end{tabular}
}
\endgroup

\addtocounter{table}{-1}                                   
\begingroup                                                
\TABLE{
{\caption{ -- continued.}}                                                          
\begin{tabular}{|c|c|c|c|c|c|r|r|r|r|r|}                   
\hline
\hline
$x$ bin&$Q^2$ bin & $\langle x \rangle$ & $\langle Q^2 \rangle\,$   
&$\displaystyle\frac{d^2\sigma^p}{dx\,dQ^2}$
 & $ F_2^p$& $\delta_{stat.}$
 & $\delta_{PID}$ & $\delta_{model}$ & $\delta_{mis.}$ & $\delta_{rad.}$ \\
       &         &                       &  [GeV$^2$]              & [nb/GeV$^2$] & 
& [\%] & [\%] & [\%] & [\%] &[\%]\\
\hline                                 
13 & E & 0.1335 & 3.538 & 0.331$\times 10^2$ & 0.356 & 0.82 & 0.02 & 0.29 & 0.91 & 0.01\\
14 & E & 0.1664 & 4.051 & 0.210$\times 10^2$ & 0.354 & 0.56 & 0.04 & 0.23 & 1.32 & 0.04\\
15 & E & 0.2107 & 4.702 & 0.122$\times 10^2$ & 0.336 & 0.58 & 0.05 & 0.18 & 1.02 & 0.05\\
16 & E & 0.2726 & 5.685 & 0.603$\times 10$ & 0.304 & 0.66 & 0.04 & 0.14 & 2.03 & 0.03\\
17 & E & 0.3679 & 9.154 & 0.124$\times 10$ & 0.234 & 0.68 & 0.05 & 0.27 & 1.40 & 0.01\\
9 & F & 0.0603 & 2.319 & 0.124$\times 10^3$ & 0.330 & 0.76 & 2.40 & 1.35 & 0.77 & 0.13\\
10 & F & 0.0734 & 2.806 & 0.726$\times 10^2$ & 0.339 & 0.69 & 2.17 & 1.01 & 0.48 & 0.15\\
11 & F & 0.0894 & 3.341 & 0.444$\times 10^2$ & 0.349 & 0.68 & 1.58 & 0.77 & 0.66 & 0.11\\
12 & F & 0.1094 & 3.946 & 0.267$\times 10^2$ & 0.346 & 0.62 & 0.96 & 0.60 & 1.03 & 0.08\\
13 & F & 0.1347 & 4.618 & 0.164$\times 10^2$ & 0.346 & 0.58 & 0.41 & 0.46 & 0.56 & 0.03\\
14 & F & 0.1679 & 5.344 & 0.102$\times 10^2$ & 0.342 & 0.58 & 0.11 & 0.34 & 1.45 & 0.04\\
15 & F & 0.2128 & 6.266 & 0.592$\times 10$ & 0.330 & 0.63 & 0.02 & 0.27 & 2.05 & 0.02\\
16 & F & 0.2759 & 7.501 & 0.299$\times 10$ & 0.296 & 0.57 & 0.04 & 0.24 & 0.85 & 0.01\\
\hline
\hline
\end{tabular}
}
\endgroup

\begingroup
\TABLE{
{\caption{\label{tab:f2d} 
Results on the differential Born cross section $\frac{d^2\sigma^d}{dx\,dQ^2}$ 
and $F_2^d$. The statistical uncertainty $\delta_{stat.}$ and 
the systematic uncertainties
$\delta_{PID}$ (particle identification), 
$\delta_{model}$ (model dependence outside the acceptance),
 $\delta_{mis.}$ (misalignment), and
$\delta_{rad.}$  (Bethe-Heitler efficiencies), 
are given in percent. Corresponding $x$ bin numbers and $Q^2$ bin numbers 
and the average values $\langle x \rangle$ and $\langle{Q^2}\rangle$  are listed in the 
first four columns. 
The overall normalization uncertainty is 7.5\%.
The structure function $F_2^d$  is derived using the parameterization 
$R=R_{1998}$~\cite{Abe:R1998}.}
}
\begin{tabular}{|c|c|c|c|c|c|r|r|r|r|r|}
\hline
\hline
$x$ bin&$Q^2$ bin& $\langle x \rangle$ & $\langle Q^2 \rangle\,$   &
 $\displaystyle\frac{d^2\sigma^d}{dx\,dQ^2}\,$ & $ F_2^d$& $\delta_{stat.}$
 & $\delta_{PID}$ & $\delta_{model}$ & $\delta_{mis.}$ & $\delta_{rad.}$ \\
       &         &                       &  [GeV$^2$]              & [nb/GeV$^2$] & & 
[\%] & [\%] & [\%] & [\%] & [\%]\\
\hline
1 & A & 0.0069 & 0.291 & 0.327$\times 10^5$ & 0.164 & 1.41 & 0.69 & 3.14 & 5.36 & 0.36\\
2 & A & 0.0102 & 0.404 & 0.138$\times 10^5$ & 0.193 & 1.30 & 0.87 & 2.15 & 3.95 & 0.47\\
3 & A & 0.0142 & 0.514 & 0.724$\times 10^4$ & 0.217 & 0.84 & 0.61 & 1.70 & 4.36 & 0.40\\
4 & A & 0.0186 & 0.626 & 0.435$\times 10^4$ & 0.242 & 0.95 & 0.36 & 1.53 & 3.89 & 0.30\\
5 & A & 0.0244 & 0.725 & 0.304$\times 10^4$ & 0.274 & 0.81 & 0.10 & 1.43 & 4.66 & 0.27\\
6 & A & 0.0325 & 0.872 & 0.178$\times 10^4$ & 0.291 & 0.87 & 0.01 & 1.38 & 3.78 & 0.22\\
7 & A & 0.0398 & 0.903 & 0.155$\times 10^4$ & 0.306 & 1.32 & 0.05 & 1.50 & 3.92 & 0.10\\
8 & A & 0.0486 & 0.929 & 0.136$\times 10^4$ & 0.323 & 2.26 & 0.07 & 2.63 & 6.43 & 0.29\\
9 & A & 0.0595 & 1.102 & 0.803$\times 10^3$ & 0.323 & 2.60 & 0.05 & 1.33 & 3.17 & 0.70\\
10 & A & 0.0724 & 1.127 & 0.722$\times 10^3$ & 0.348 & 2.62 & 0.06 & 1.14 & 5.49 & 0.45\\
11 & A & 0.0882 & 1.167 & 0.564$\times 10^3$ & 0.338 & 2.53 & 0.07 & 0.84 & 3.45 & 1.18\\
12 & A & 0.1078 & 1.205 & 0.460$\times 10^3$ & 0.345 & 3.66 & 0.07 & 0.75 & 5.93 & 1.18\\
13 & A & 0.1330 & 1.263 & 0.335$\times 10^3$ & 0.329 & 3.06 & 0.07 & 0.78 & 2.56 & 0.94\\
14 & A & 0.1657 & 1.337 & 0.265$\times 10^3$ & 0.354 & 2.65 & 0.06 & 1.02 & 7.52 & 0.90\\
15 & A & 0.2093 & 1.447 & 0.155$\times 10^3$ & 0.298 & 2.21 & 0.06 & 1.87 & 2.99 & 0.52\\
\hline
\hline
\end{tabular}                                             
}
\endgroup

\addtocounter{table}{-1}
\begingroup
\TABLE{
{\caption{ -- continued.}
}
\begin{tabular}{|c|c|c|c|c|c|r|r|r|r|r|}
\hline
\hline
$x$ bin&$Q^2$ bin& $\langle x \rangle$ & $\langle Q^2 \rangle\,$  
& $\displaystyle\frac{d^2\sigma^d}{dx\,dQ^2}\, $ & $ F_2^d$& $\delta_{stat.}$ &
$\delta_{PID}$ &$\delta_{model}$ & $\delta_{mis.}$ & $\delta_{rad.}$ \\
       &         &                       &  [GeV$^2$]              
& [nb/GeV$^2$] & & [\%] & [\%] & [\%] & [\%] &[\%] \\
\hline
16 & A & 0.2642 & 1.749 & 0.845$\times 10^2$ & 0.298 & 1.25 & 0.05 & 2.65 & 3.29 & 0.56\\
17 & A & 0.3477 & 2.672 & 0.196$\times 10^2$ & 0.217 & 1.37 & 0.05 & 1.99 & 7.62 & 0.21\\
18 & A & 0.4605 & 4.224 & 0.377$\times 10$ & 0.142 & 1.13 & 0.04 & 1.70 & 4.10 & 0.11\\
19 & A & 0.6346 & 8.688 & 0.198 & 0.047 & 2.80 & 0.08 & 1.25 & 1.17 & 0.02\\
1 & B & 0.0085 & 0.354 & 0.203$\times 10^5$ & 0.189 & 1.29 & 0.93 & 2.43 & 3.75 & 0.81\\
2 & B & 0.0117 & 0.477 & 0.979$\times 10^4$ & 0.226 & 0.77 & 1.22 & 2.05 & 2.07 & 0.69\\
3 & B & 0.0152 & 0.612 & 0.488$\times 10^4$ & 0.241 & 0.61 & 1.50 & 1.90 & 1.03 & 0.36\\
4 & B & 0.0193 & 0.770 & 0.260$\times 10^4$ & 0.258 & 0.64 & 1.80 & 1.79 & 0.79 & 0.40\\
5 & B & 0.0250 & 0.896 & 0.169$\times 10^4$ & 0.272 & 0.52 & 1.07 & 1.51 & 0.56 & 0.12\\
6 & B & 0.0327 & 1.144 & 0.865$\times 10^3$ & 0.291 & 0.68 & 0.92 & 1.47 & 0.64 & 0.30\\
7 & B & 0.0401 & 1.074 & 0.100$\times 10^4$ & 0.308 & 1.06 & 0.01 & 1.28 & 2.61 & 0.17\\
8 & B & 0.0490 & 1.123 & 0.843$\times 10^3$ & 0.318 & 1.05 & 0.03 & 1.22 & 2.47 & 0.24\\
9 & B & 0.0596 & 1.281 & 0.582$\times 10^3$ & 0.337 & 2.05 & 0.06 & 0.81 & 2.26 & 0.36\\
10 & B & 0.0725 & 1.359 & 0.439$\times 10^3$ & 0.329 & 1.85 & 0.06 & 0.58 & 3.96 & 0.36\\
11 & B & 0.0884 & 1.399 & 0.383$\times 10^3$ & 0.349 & 1.96 & 0.06 & 0.51 & 3.58 & 0.05\\
12 & B & 0.1082 & 1.554 & 0.260$\times 10^3$ & 0.348 & 1.22 & 0.06 & 0.47 & 4.34 & 0.19\\
13 & B & 0.1333 & 1.632 & 0.195$\times 10^3$ & 0.339 & 1.85 & 0.06 & 0.49 & 1.84 & 0.34\\
14 & B & 0.1661 & 1.811 & 0.132$\times 10^3$ & 0.341 & 1.01 & 0.05 & 0.50 & 4.10 & 0.05\\
15 & B & 0.2104 & 2.050 & 0.763$\times 10^2$ & 0.314 & 1.53 & 0.05 & 0.58 & 2.80 & 0.09\\
16 & B & 0.2722 & 2.471 & 0.374$\times 10^{2}$ & 0.285 & 1.02 & 0.05 & 0.78 & 3.85 & 0.18\\
17 & B & 0.3620 & 3.790 & 0.893$\times 10$ & 0.218 & 0.86 & 0.04 & 0.57 & 2.21 & 0.04\\
18 & B & 0.4900 & 6.026 & 0.139$\times 10$ & 0.120 & 0.65 & 0.04 & 0.62 & 0.94 & 0.01\\
19 & B & 0.6641 & 12.778 & 0.605$\times 10^{-1}$ & 0.036 & 2.32 & 0.05 & 0.66 & 0.86 & 0.01\\
5 & C & 0.0270 & 1.092 & 0.103$\times 10^4$ & 0.292 & 0.68 & 2.39 & 1.81 & 1.06 & 0.25\\
6 & C & 0.0338 & 1.391 & 0.535$\times 10^3$ & 0.309 & 0.99 & 2.98 & 1.93 & 0.28 & 0.43\\
7 & C & 0.0401 & 1.293 & 0.621$\times 10^3$ & 0.309 & 0.80 & 0.38 & 1.35 & 2.71 & 0.15\\
8 & C & 0.0490 & 1.460 & 0.430$\times 10^3$ & 0.316 & 0.70 & 0.17 & 1.05 & 1.29 & 0.12\\
9 & C & 0.0596 & 1.466 & 0.400$\times 10^3$ & 0.323 & 1.09 & 0.01 & 0.69 & 1.73 & 0.24\\
10 & C & 0.0726 & 1.627 & 0.291$\times 10^3$ & 0.337 & 1.10 & 0.04 & 0.48 & 2.05 & 0.07\\
11 & C & 0.0885 & 1.691 & 0.236$\times 10^3$ & 0.337 & 1.11 & 0.05 & 0.41 & 1.43 & 0.02\\
12 & C & 0.1082 & 2.013 & 0.140$\times 10^3$ & 0.341 & 0.96 & 0.05 & 0.37 & 2.70 & 0.08\\
13 & C & 0.1334 & 2.158 & 0.102$\times 10^3$ & 0.335 & 1.09 & 0.05 & 0.33 & 2.17 & 0.05\\
14 & C & 0.1662 & 2.458 & 0.636$\times 10^2$ & 0.328 & 1.22 & 0.05 & 0.30 & 0.98 & 0.07\\
15 & C & 0.2105 & 2.851 & 0.361$\times 10^2$ & 0.310 & 0.75 & 0.04 & 0.29 & 2.40 & 0.07\\
16 & C & 0.2724 & 3.479 & 0.167$\times 10^2$ & 0.272 & 0.67 & 0.04 & 0.28 & 1.74 & 0.03\\
17 & C & 0.3622 & 5.125 & 0.428$\times 10$ & 0.205 & 0.62 & 0.04 & 0.21 & 1.84 & 0.04\\
18 & C & 0.4957 & 8.280 & 0.630 & 0.113 & 0.70 & 0.05 & 0.22 & 2.23 & 0.02\\
\hline
\hline
\end{tabular}                                             
}
\endgroup

\addtocounter{table}{-1}
\begingroup
\TABLE{
{\caption{ -- continued.}
}
\begin{tabular}{|c|c|c|c|c|c|r|r|r|r|r|}
\hline
\hline
$x$ bin&$Q^2$ bin& $\langle x \rangle$ & $\langle Q^2 \rangle\,$  
& $\displaystyle\frac{d^2\sigma^d}{dx\,dQ^2}\, $ & $ F_2^d$& $\delta_{stat.}$ & 
 $\delta_{PID}$ & $\delta_{model}$ &$\delta_{mis.}$ & $\delta_{rad.}$ \\
       &         &                       &  [GeV$^2$]              
& [nb/GeV$^2$] & & [\%] & [\%] & [\%] & [\%] & [\%]\\
\hline 
7 & D & 0.0408 & 1.615 & 0.343$\times 10^3$ & 0.311 & 0.68 & 2.58 & 1.76 & 0.53 & 0.34\\
8 & D & 0.0496 & 1.927 & 0.208$\times 10^3$ & 0.320 & 0.60 & 2.49 & 1.49 & 0.68 & 0.10\\
9 & D & 0.0596 & 1.675 & 0.290$\times 10^3$ & 0.328 & 1.21 & 0.04 & 0.69 & 1.88 & 0.06\\
10 & D & 0.0726 & 1.928 & 0.192$\times 10^3$ & 0.339 & 1.46 & 0.01 & 0.46 & 2.13 & 0.01\\
11 & D & 0.0885 & 2.080 & 0.144$\times 10^3$ & 0.339 & 0.89 & 0.03 & 0.37 & 1.21 & 0.07\\
12 & D & 0.1083 & 2.490 & 0.824$\times 10^2$ & 0.336 & 1.01 & 0.04 & 0.31 & 1.50 & 0.18\\
13 & D & 0.1334 & 2.795 & 0.570$\times 10^2$ & 0.345 & 0.83 & 0.04 & 0.26 & 2.67 & 0.11\\
14 & D & 0.1663 & 3.195 & 0.343$\times 10^2$ & 0.326 & 0.90 & 0.04 & 0.22 & 1.91 & 0.01\\
15 & D & 0.2106 & 3.709 & 0.195$\times 10^2$ & 0.305 & 0.69 & 0.05 & 0.19 & 1.80 & 0.04\\
16 & D & 0.2726 & 4.527 & 0.899$\times 10$ & 0.266 & 0.88 & 0.05 & 0.17 & 2.14 & 0.04\\
17 & D & 0.3624 & 6.797 & 0.215$\times 10$ & 0.197 & 0.58 & 0.04 & 0.12 & 1.63 & 0.01\\
18 & D & 0.5038 & 11.344 & 0.272 & 0.103 & 1.03 & 0.07 & 0.31 & 1.61 & 0.02\\
9 & E & 0.0597 & 1.906 & 0.202$\times 10^3$ & 0.321 & 0.90 & 0.30 & 0.74 & 0.78 & 0.08\\
10 & E & 0.0726 & 2.276 & 0.124$\times 10^3$ & 0.334 & 0.92 & 0.22 & 0.53 & 1.53 & 0.13\\
11 & E & 0.0885 & 2.587 & 0.826$\times 10^2$ & 0.335 & 0.68 & 0.09 & 0.40 & 1.33 & 0.11\\
12 & E & 0.1083 & 3.071 & 0.494$\times 10^2$ & 0.338 & 0.55 & 0.03 & 0.32 & 1.49 & 0.03\\
13 & E & 0.1335 & 3.538 & 0.306$\times 10^2$ & 0.330 & 0.82 & 0.02 & 0.25 & 0.91 & 0.04\\
14 & E & 0.1664 & 4.051 & 0.191$\times 10^2$ & 0.321 & 0.54 & 0.04 & 0.19 & 1.32 & 0.03\\
15 & E & 0.2107 & 4.702 & 0.108$\times 10^2$ & 0.298 & 0.57 & 0.05 & 0.15 & 1.02 & 0.05\\
16 & E & 0.2726 & 5.685 & 0.520$\times 10$ & 0.262 & 0.66 & 0.04 & 0.12 & 2.03 & 0.00\\
17 & E & 0.3679 & 9.154 & 0.101$\times 10$ & 0.189 & 0.69 & 0.05 & 0.25 & 1.40 & 0.02\\
9 & F & 0.0603 & 2.319 & 0.122$\times 10^3$ & 0.324 & 0.73 & 2.40 & 1.15 & 0.77 & 0.06\\
10 & F & 0.0734 & 2.806 & 0.705$\times 10^2$ & 0.329 & 0.65 & 2.17 & 0.87 & 0.48 & 0.03\\
11 & F & 0.0894 & 3.341 & 0.421$\times 10^2$ & 0.331 & 0.65 & 1.58 & 0.64 & 0.66 & 0.07\\
12 & F & 0.1094 & 3.946 & 0.255$\times 10^2$ & 0.331 & 0.58 & 0.96 & 0.51 & 1.03 & 0.07\\
13 & F & 0.1347 & 4.618 & 0.153$\times 10^2$ & 0.322 & 0.55 & 0.41 & 0.40 & 0.56 & 0.06\\
14 & F & 0.1679 & 5.344 & 0.930$\times 10$ & 0.311 & 0.56 & 0.11 & 0.30 & 1.45 & 0.02\\
15 & F & 0.2128 & 6.266 & 0.524$\times 10$ & 0.292 & 0.63 & 0.02 & 0.24 & 2.05 & 0.02\\
16 & F & 0.2759 & 7.501 & 0.251$\times 10$ & 0.249 & 0.57 & 0.04 & 0.22 & 0.85 & 0.03\\
\hline
\hline
\end{tabular}                                                                       
}
\endgroup

\begingroup
\TABLE{
{\caption{\label{tab:f2ratio} 
Results on the inelastic Born cross-section ratio ${\sigma^d}/{\sigma^p}$. 
The statistical uncertainty $\delta_{stat.}$,
the systematic uncertainty $\delta_{rad.}$ due to radiative corrections
and $\delta_{model}$ due to the model dependence outside the acceptance
are given in percent. Corresponding $x$ bin numbers and $Q^2$ bin numbers 
and the average values of $x$ and $Q^2$ are listed in the 
first four columns. The overall normalization uncertainty is 1.4\,$\%$.}
}
\begin{tabular}[t]{|c|c|c|c|c|c|c|c|}
\hline
\hline
$x$ bin & $Q^2$ bin & $\langle x \rangle$ & $\langle Q^2 \rangle$ & 
$\displaystyle\sigma^d/\sigma^p$  &  $\delta_{stat.}$ & $\delta_{rad.}$&$\delta_{model}$\\
       &           &         &  [GeV$^2$] & & [\%]& [\%]& [\%]\\
\hline
1 & A & 0.0069 & 0.291 & 0.992 & 2.19 & 1.52 & 3.85 \\
2 & A & 0.0102 & 0.404 & 0.952 & 1.93 & 0.94 & 2.53 \\
3 & A & 0.0142 & 0.514 & 0.979 & 1.27 & 0.43 & 2.01 \\
4 & A & 0.0186 & 0.626 & 0.984 & 1.41 & 0.39 & 1.71 \\
5 & A & 0.0244 & 0.725 & 0.977 & 1.21 & 0.10 & 1.56 \\
6 & A & 0.0325 & 0.872 & 0.965 & 1.26 & 0.57 & 1.47 \\
7 & A & 0.0398 & 0.903 & 0.972 & 1.94 & 0.82 & 1.62 \\
8 & A & 0.0486 & 0.929 & 0.962 & 3.24 & 0.54 & 2.70 \\
9 & A & 0.0595 & 1.102 & 0.948 & 3.72 & 1.49 & 1.37 \\
10 & A & 0.0724 & 1.127 & 0.936 & 3.69 & 0.63 & 1.17 \\
11 & A & 0.0882 & 1.167 & 0.954 & 3.51 & 1.84 & 0.89 \\
12 & A & 0.1078 & 1.205 & 0.923 & 5.02 & 1.51 & 0.81 \\
13 & A & 0.1330 & 1.263 & 0.885 & 4.14 & 1.97 & 0.86 \\
14 & A & 0.1657 & 1.337 & 0.927 & 3.57 & 2.48 & 1.09 \\
15 & A & 0.2093 & 1.447 & 0.873 & 2.87 & 1.24 & 1.90 \\
16 & A & 0.2642 & 1.749 & 0.848 & 1.72 & 1.09 & 2.76 \\
17 & A & 0.3477 & 2.672 & 0.798 & 1.88 & 0.09 & 2.08 \\
18 & A & 0.4605 & 4.224 & 0.783 & 1.59 & 0.19 & 1.77 \\
19 & A & 0.6346 & 8.688 & 0.743 & 3.90 & 0.07 & 1.28 \\
1 & B & 0.0085 & 0.354 & 0.968 & 1.97 & 0.69 & 2.97 \\
2 & B & 0.0117 & 0.477 & 0.967 & 1.17 & 0.99 & 2.49 \\
3 & B & 0.0152 & 0.612 & 0.972 & 0.93 & 0.81 & 2.28 \\
4 & B & 0.0193 & 0.770 & 0.969 & 0.97 & 0.33 & 2.07 \\
5 & B & 0.0250 & 0.896 & 0.981 & 0.78 & 0.27 & 1.69 \\
6 & B & 0.0327 & 1.144 & 0.972 & 1.00 & 0.45 & 1.60 \\
7 & B & 0.0401 & 1.074 & 0.964 & 1.56 & 0.32 & 1.38 \\
8 & B & 0.0490 & 1.123 & 0.962 & 1.55 & 0.07 & 1.30 \\
9 & B & 0.0596 & 1.281 & 0.969 & 2.96 & 0.57 & 0.87 \\
10 & B & 0.0725 & 1.359 & 0.928 & 2.63 & 0.18 & 0.61 \\
11 & B & 0.0884 & 1.399 & 0.925 & 2.71 & 0.14 & 0.55 \\
12 & B & 0.1082 & 1.554 & 0.921 & 1.73 & 0.36 & 0.50 \\
13 & B & 0.1333 & 1.632 & 0.901 & 2.57 & 0.73 & 0.53 \\
14 & B & 0.1661 & 1.811 & 0.903 & 1.45 & 0.17 & 0.54 \\
15 & B & 0.2104 & 2.050 & 0.885 & 2.13 & 0.25 & 0.61 \\
16 & B & 0.2722 & 2.471 & 0.860 & 1.43 & 0.06 & 0.82 \\
17 & B & 0.3620 & 3.790 & 0.809 & 1.21 & 0.09 & 0.60 \\
18 & B & 0.4900 & 6.026 & 0.768 & 0.91 & 0.03 & 0.66 \\
19 & B & 0.6641 & 12.778 & 0.756 & 3.22 & 0.09 & 0.67 \\
5 & C & 0.0270 & 1.092 & 0.990 & 1.04 & 0.43 & 2.03 \\
\hline
\hline
\end{tabular}                      
}
\endgroup

\addtocounter{table}{-1}
\begingroup
\TABLE{
{\caption{ -- continued.}}
\begin{tabular}[t]{|c|c|c|c|c|c|c|c|}
\hline
\hline
$x$ bin & $Q^2$ bin & $\langle x \rangle$ & $\langle Q^2 \rangle$ 
& $\sigma^d/\sigma^p$  &  $\delta_{stat.}$ & $\delta_{rad.}$&$\delta_{model}$\\
       &           &         &  [GeV$^2$] & & [\%]& [\%]& [\%]\\
\hline
6 & C & 0.0338 & 1.391 & 0.990 & 1.49 & 0.73 & 2.15 \\
7 & C & 0.0401 & 1.293 & 0.975 & 1.18 & 0.07 & 1.45 \\
8 & C & 0.0490 & 1.460 & 0.952 & 1.02 & 0.19 & 1.12 \\
9 & C & 0.0596 & 1.466 & 0.941 & 1.59 & 0.20 & 0.75 \\
10 & C & 0.0726 & 1.627 & 0.943 & 1.59 & 0.26 & 0.53 \\
11 & C & 0.0885 & 1.691 & 0.945 & 1.60 & 0.08 & 0.46 \\
12 & C & 0.1082 & 2.013 & 0.917 & 1.39 & 0.05 & 0.39 \\
13 & C & 0.1334 & 2.158 & 0.920 & 1.57 & 0.17 & 0.36 \\
14 & C & 0.1662 & 2.458 & 0.891 & 1.74 & 0.10 & 0.33 \\
15 & C & 0.2105 & 2.851 & 0.862 & 1.07 & 0.01 & 0.31 \\
16 & C & 0.2724 & 3.479 & 0.851 & 0.96 & 0.08 & 0.30 \\
17 & C & 0.3622 & 5.125 & 0.813 & 0.87 & 0.03 & 0.22 \\
18 & C & 0.4957 & 8.280 & 0.760 & 0.97 & 0.02 & 0.23 \\
7 & D & 0.0408 & 1.615 & 0.969 & 1.01 & 0.51 & 1.94 \\
8 & D & 0.0496 & 1.927 & 0.978 & 0.90 & 0.24 & 1.63 \\
9 & D & 0.0596 & 1.675 & 0.961 & 1.76 & 0.22 & 0.74 \\
10 & D & 0.0726 & 1.928 & 0.951 & 2.11 & 0.12 & 0.51 \\
11 & D & 0.0885 & 2.080 & 0.946 & 1.31 & 0.10 & 0.43 \\
12 & D & 0.1083 & 2.490 & 0.914 & 1.46 & 0.31 & 0.34 \\
13 & D & 0.1334 & 2.795 & 0.923 & 1.20 & 0.21 & 0.29 \\
14 & D & 0.1663 & 3.195 & 0.896 & 1.29 & 0.10 & 0.25 \\
15 & D & 0.2106 & 3.709 & 0.877 & 0.99 & 0.10 & 0.21 \\
16 & D & 0.2726 & 4.527 & 0.834 & 1.25 & 0.02 & 0.18 \\
17 & D & 0.3624 & 6.797 & 0.808 & 0.82 & 0.02 & 0.13 \\
18 & D & 0.5038 & 11.344 & 0.770 & 1.43 & 0.02 & 0.31 \\
9 & E & 0.0597 & 1.906 & 0.950 & 1.31 & 0.18 & 0.81 \\
10 & E & 0.0726 & 2.276 & 0.954 & 1.34 & 0.20 & 0.58 \\
11 & E & 0.0885 & 2.587 & 0.944 & 1.00 & 0.14 & 0.44 \\
12 & E & 0.1083 & 3.071 & 0.940 & 0.81 & 0.04 & 0.35 \\
13 & E & 0.1335 & 3.538 & 0.912 & 1.20 & 0.04 & 0.27 \\
14 & E & 0.1664 & 4.051 & 0.901 & 0.79 & 0.06 & 0.21 \\
15 & E & 0.2107 & 4.702 & 0.878 & 0.82 & 0.04 & 0.17 \\
16 & E & 0.2726 & 5.685 & 0.853 & 0.95 & 0.04 & 0.13 \\
17 & E & 0.3679 & 9.154 & 0.802 & 0.98 & 0.02 & 0.26 \\
9 & F & 0.0603 & 2.319 & 0.969 & 1.07 & 0.14 & 1.26 \\
10 & F & 0.0734 & 2.806 & 0.955 & 0.96 & 0.15 & 0.94 \\
11 & F & 0.0894 & 3.341 & 0.936 & 0.96 & 0.09 & 0.72 \\
12 & F & 0.1094 & 3.946 & 0.947 & 0.85 & 0.08 & 0.56 \\
13 & F & 0.1347 & 4.618 & 0.920 & 0.81 & 0.05 & 0.44 \\
14 & F & 0.1679 & 5.344 & 0.898 & 0.81 & 0.03 & 0.32 \\
15 & F & 0.2128 & 6.266 & 0.874 & 0.91 & 0.01 & 0.26 \\
16 & F & 0.2759 & 7.501 & 0.829 & 0.82 & 0.04 & 0.23 \\
\hline
\hline
\end{tabular}    
}                                                                   
\endgroup


\begingroup
\TABLE{
{  \caption{Results of fits to the cross section $\sigma_{L+T}$ 
    using the ALLM functional form~\cite{allm:1991}. Parameters of the
    ALLM97 fit ~\cite{allm:1997} and the GD07-P fit~\cite{gd07} (both
    based on proton data) are quoted  
    together with the new parameters determined from current world data 
    including the HERMES results discussed in this paper 
    on proton (GD11-P) and deuteron (GD11-D) cross sections. 
     In Ref.~\cite{gd07} the uncertainties were calculated with an UP value~\cite{minuit} 
equal to 24.7. 
That identical fit was redone here with the same value of UP=1
as the new fits in order to provide an appropriate comparison of its uncertainties. 
    The ALLM97 fit provided no uncertainties.
    The uncertainties given in the table for GD07-P, GD11-P, and GD11-D 
    correspond to the diagonal elements of the full covariance matrix
    which must be used to calculate uncertainties in $F_2$ or cross sections. 
They can be obtained by contacting the management of the HERMES 
collaboration: $<$management@hermes.desy.de$>$.
    The parameter $\Lambda^2_0$ has no uncertainty as it is fixed in the fits. 
In the case of the deuteron fit, the parameters $m^2_0$ and $Q^2_0$ had to be fixed   
in order to obtain a good fit.
    \label{tab:parameters}
  }}
\begin{tabular}[t]{|c||c|cc|cc|cc|}
\hline
\hline
Parameter &  ALLM97-P  & \multicolumn{2}{c|}{GD07-P}  & \multicolumn{2}{c|}{GD11-P}  & \multicolumn{2}{c|}{GD11-D} \\
           &  value    &  value & unc.                & value & unc.                 & value & unc.          \\ 
\hline
$m_0^2 [$GeV$^2]$           & 0.31985 & 0.454   & 0.0283    &   0.5063    &   0.0236   &   0.426    &     -   \\
$m_\mathcal{P}^2 [$GeV$^2]$  & 49.457  & 30.7    & 2.85      &   34.75     &   2.56     & 0.00007713 & 0.000319\\
$m_\mathcal{R}^2 [$GeV$^2]$  & 0.15052 & 0.117   & 0.0465    &   0.03190   &   0.00828  & 0.2293     & 0.0763  \\
$Q_0^2 [$GeV$^2]$           & 0.52544 & 1.15    & 0.358     &   1.374     &   0.349    & 2.65       &    -    \\
$\Lambda_0^2 [$GeV$^2]$     & 0.06527 & 0.06527 & -         &   0.06527   &   -        & 0.06527    &    -    \\
$a_{{\mathcal{P}}1}$         & -0.0808 & -0.105  & 0.00507   &  -0.11895    &   0.00411  &-0.4287     & 0.0773  \\
$a_{{\mathcal{P}}2}$         & 0.44812 & -0.495  & 0.0306    &  -0.4783     &   0.0224   & 0.2891     & 0.0646  \\
$a_{{\mathcal{P}}3}$         & 1.1709  & 1.29    & 0.243     &   1.353      &   0.203    & 0.3931     & 0.0935  \\
$b_{{\mathcal{P}}4}$         & 0.36292 & -1.42   & 0.491     &   1.0833     &   0.0908   &-27.212     & 0.649   \\
$b_{{\mathcal{P}}5}$         & 1.8917  & 4.51    & 0.540     &   2.656      &   0.363    &30.687      & 0.650   \\
$b_{{\mathcal{P}}6}$         & 1.8439  & 0.551   & 0.120     &   1.771      &   0.204    & 0.04577    & 0.00198 \\
$c_{{\mathcal{P}}7}$         & 0.28067 & 0.339   & 0.0194    &   0.3638     &   0.0140   & 0.00073    & 0.00181 \\
$c_{{\mathcal{P}}8}$         & 0.22291 & 0.127   & 0.0217    &   0.1211     &   0.0163   & 0.9741     & 0.0566  \\
$c_{{\mathcal{P}}9}$         & 2.1979  & 1.16    & 0.246     &   1.166      &   0.197    & 0.8722     & 0.0800  \\
$a_{{\mathcal{R}}1}$         & 0.584   & 0.374   & 0.0320    &   0.3425     &   0.0243   & 0.2986     & 0.0237  \\
$a_{{\mathcal{R}}2}$         & 0.37888 & 0.998   & 0.102     &   1.0603     &   0.0640   & 3.615      & 0.230   \\
$a_{{\mathcal{R}}3}$         & 2.6063  & 0.775   & 0.112     &   0.5164     &   0.0440   & 1.1455     & 0.0653  \\
$b_{{\mathcal{R}}4}$         & 0.01147 & 2.71    & 0.393     &   -10.408    &   0.627    & 1.987      & 0.333   \\
$b_{{\mathcal{R}}5}$         & 3.7582  & 1.83    & 0.537     &   14.857     &   0.627    & 7.150      & 0.343   \\
$b_{{\mathcal{R}}6}$         & 0.49338 & 1.26    & 0.296     &   0.07739    &   0.00861  & 0.9350     & 0.0935  \\
$c_{{\mathcal{R}}7}$         & 0.80107 & 0.838   & 0.106     &   1.3633     &   0.0867   & 1.0316     & 0.0783  \\
$c_{{\mathcal{R}}8}$         & 0.97307 & 2.36    & 0.557     &   2.256      &   0.503    & 26.36      & 6.19    \\
$c_{{\mathcal{R}}9}$         & 3.4942  & 1.77    & 0.209     &   2.209      &   0.260    &  3.024     & 0.136   \\
\hline
\hline
\end{tabular}
}
\endgroup

\begingroup
\TABLE{
{  \caption{Normalizations $\delta_k\nu_k$ (see Eq.~(\ref{Eq:chi2})) 
obtained in the GD11 fits.
    \label{tab:gd10normalizations}
 }}
\begin{tabular}[t]{|l|c|c|}
\hline
\hline
Data set &  GD11-P & GD11-D  \\
\hline
HERA (positron beam)~\cite{:2009wt}          & -0.0065  &  -      \\
HERA (electron beam)~\cite{:2009wt}          & -0.0067  &   -     \\
E665~\cite{e665}                             &  0.020   & -0.014  \\
NMC-90~GeV~\cite{nmc9610231}                 & -0.00020 & -0.029  \\
NMC-120~GeV~\cite{nmc9610231}                &  0.011   & -0.0096 \\
NMC-200~GeV~\cite{nmc9610231}                &  0.0093  &  0.0036 \\
NMC-280~GeV~\cite{nmc9610231}                &  0.0035  &  0.0023 \\
BCDMS-100~GeV~\cite{bcdms}                   & -0.032   &  -      \\
BCDMS-120~GeV~\cite{bcdms}                   & -0.028   & -0.0075 \\
BCDMS-200~GeV~\cite{bcdms}                   & -0.027   & -0.0060 \\
BCDMS-280~GeV~\cite{bcdms}                   & -0.023   & -0.0031 \\
SLAC E49a~\cite{whitlow:thesis}              &  0.016   & -0.0013 \\
SLAC E49b~\cite{whitlow:thesis}              &  0.022   &  0.0062 \\
SLAC E61~\cite{whitlow:thesis}               &  0.016   &  0.0070 \\
SLAC E87~\cite{whitlow:thesis}               &  0.016   &  0.0045 \\
SLAC E89a~\cite{whitlow:thesis}              &  0.036   &  0.0087 \\ 
SLAC E89b~\cite{whitlow:thesis}              &  0.018   &  0.00081\\
SLAC E139~\cite{whitlow:thesis}              &     -    &  0.0014 \\
SLAC E140~\cite{whitlow:thesis}              &     -    &  0.0025 \\
JLAB~\cite{Malace:2009kw}                    & -0.012   & -0.040  \\
JLAB~\cite{Osipenko:2003ua}                  & -0.0063  &  -      \\
JLAB~\cite{Osipenko:2005gt}                  &  -       &  -0.0012\\
JLAB (Rosenbluth)~\cite{Tvaskis:2010as}      &  0.0014  &  0.0088 \\
JLAB (Model Dependent)~\cite{Tvaskis:2010as} &  0.0085  &  0.0088 \\
HERMES (T.A.)                                &  0.015   & -0.022  \\
\hline
\hline
\end{tabular}
}
\endgroup


\end{appendix}
\end{document}